\documentclass[aps,pra,twocolumn,superscriptaddress,floatfix]{revtex4-1}
\usepackage{color}
\usepackage{amsmath,amsfonts,amssymb}
\usepackage{amsthm}
\usepackage{bm}
\usepackage{graphicx}
\usepackage{bbm}
\usepackage{epstopdf}
\usepackage{epsfig}
\usepackage{verbatim}
\usepackage{array}
\usepackage{ulem}
\usepackage{notes2bib}
\usepackage[T1]{fontenc}
\usepackage{tgtermes}

\newcommand{\appsection}[1]{\section{\uppercase{#1}}}

\DeclareMathOperator{\calD}{\mathcal{D}}
\DeclareMathOperator{\calZ}{\mathcal{Z}}
\DeclareMathOperator{\ketzd}{| \empty \rangle |\!\cdot\!\cdot \rangle}
\DeclareMathOperator{\ketoo}{| \cdot \rangle |\cdot \rangle}
\DeclareMathOperator{\ketdz}{| \!\cdot\!\cdot \rangle |\empty \rangle}

\usepackage[unicode=true,bookmarks=true,bookmarksnumbered=false,bookmarksopen=false, breaklinks=true,pdfborder={0 0 1},backref=false,colorlinks=true]{hyperref}
\hypersetup{linkcolor=magenta, urlcolor=blue, citecolor=blue, pdfstartview={FitH}, unicode=true}

\begin{document}
\title{Nonzero angular momentum density wave phases in SU($N$) fermions with
singlet-bond and triplet-current interactions}
\author{Han Xu}
\affiliation{School of Physics and Technology, Wuhan University, Wuhan 430072, China}
\affiliation{Department of Physics, City University of Hong Kong, Tat Chee Avenue, Kowloon, Hong Kong SAR, China, and City University of Hong Kong Shenzhen Research Institute, Shenzhen, Guangdong 518057, China}
\author{Congjun Wu}
\affiliation{Department of Physics, School of Science, Westlake University, Hangzhou 310024, Zhejiang, China}
\affiliation{Institute for Theoretical Sciences, Westlake University, Hangzhou 310024, Zhejiang, China}
\affiliation{Key Laboratory for Quantum Materials of Zhejiang Province, School of Science, Westlake University, Hangzhou 310024, China}
\affiliation{Institute of Natural Sciences, Westlake Institute for Advanced Study, Hangzhou 310024, Zhejiang, China}
\author{Yu Wang}
\email{yu.wang@whu.edu.cn}
\affiliation{School of Physics and Technology, Wuhan University, Wuhan 430072, China}

\begin{abstract}
    We employ the sign-problem-free projector determinant quantum Monte Carlo method to study a microscopic model of SU($N$) fermions with singlet-bond and triplet-current interactions on the square lattice.
    We find the gapped singlet $p_x$ and gapless triplet $d_{x^2-y^2}$ density wave states in the half-filled $N=4$ model.
    Specifically, the triplet $d_{x^2-y^2}$ density wave order is observed in the weak triplet-current interaction regime.
    As the triplet-current interaction strength is further increased, our simulations demonstrate a transition to the singlet $p_x$ density wave state, accompanied by a gapped mixed-ordered area where the two orders coexist. 
    With increasing the singlet-bond interaction strength, the triplet $d_{x^2-y^2}$-wave order persists up to a critical point after which the singlet $p_x$ density wave state is stabilized, while the ground state is disordered in between the two ordered phases.
    The analytical continuation is then performed to derive the single-particle spectrum.
    In the spectra of triplet $d_{x^2-y^2}$ and singlet $p_x$ density waves,
    the anisotropic Dirac cone and the parabolic shape around the Dirac point are observed, respectively.
    As for the mixed-ordered area, a single-particle gap opens and the velocities remain anisotropic at the Dirac point.
\end{abstract}

\maketitle

\section{Introduction}

The nonzero angular momentum density wave state is classified as the condensation of particle-hole pairs with nonzero angular momentum \cite{nayak2000density}, in analogy with the higher angular momentum superconducting state \cite{lee2006doping}, which generalizes the conventional charge density wave. 
For example, 
on the square lattice the singlet $p_x$ density wave is known as the spin dimerized state or bond-centered charge density wave in the literatures \cite{affleck1988largen,affleck1989largen}. 
For the commensurate ordering at wavevector $(\pi,0)$, the singlet $p_x$ density wave state breaks the translational and rotational symmetries, but the time-reversal and spin rotational symmetries are preserved. 
Another example is the singlet $d_{x^2-y^2}$ density wave that has a checkerboard pattern of currents around elementary plaquettes, also known as the staggered flux state \cite{affleck1988largen,affleck1989largen,wang1990flux}. 
Such state breaks the translational, rotational and time-reversal symmetries.
Aside from the singlet analogs, the triplet version of the $d_{x^2-y^2}$ density wave state has been proposed \cite{nayak2000density}, 
which is expected as the origin of the pseudogap regime in the cuprate superconductors \cite{liu2003spin,chakravarty2001hidden,maki2007d}.
Although the spin-rotational invariance is broken, the triplet $d_{x^2-y^2}$ density wave state does not have magnetic order;
meanwhile, since the spin currents are time-reversal even, it preserves the time-reversal symmetry.
On the other hand, the spin current circulates around each plaquette in an alternating pattern, 
so the translational and rotational symmetries are still broken. 
In addition, the triplet $d$-wave order on the hexagonal lattices can be defined in similar ways \cite{maharaj2013particle,venderbos2016symmetry,venderbos2016multi}.

In recent years, the unbiased and nonperturbative quantum Monte Carlo (QMC) methods have been applied to systematically explore the $p$ and $d$ density waves in the context of strongly correlated fermion systems. 
Considering the large-$N$ theories, the SU($N$) generalization of the SU(2) lattice fermion model is of particular importance
because the $p$- and $d$-wave phases are usually stabilized at large values of $N$ \cite{affleck1988largen,affleck1989largen}. 
For example, a determinant QMC study found that the singlet $p_x$ and $d_{x^2-y^2}$ density waves are the possible ground states of the SU($N$) Hubbard-Heisenberg model on the square lattice when $N\geqslant6$ \cite{assaad2005phase}. 
As for the honeycomb lattice, various spin dimerized states are stabilized when $N\geqslant4$  \cite{lang2013dimerized}. 
Also, the SU($N$) generalization can actually be implemented in the state-of-art cold-atom experiments with
large-spin alkaline-earth fermions \cite{Wu2003Exact,wu2006hidden,DeSalvo2010Degenerate,zhang2014spectroscopic,taie2010realization,taie2012ansu6,taie2022observation}.

\begin{figure}
  \includegraphics[width=0.9\linewidth]{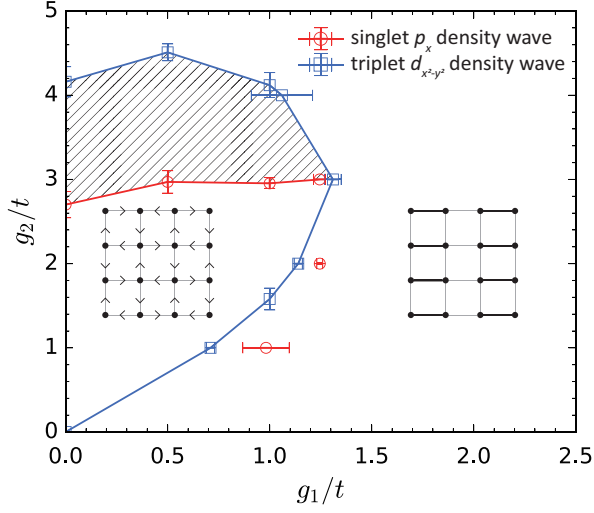}
  \caption{
      Phase diagram of the half-filled $N=4$ model Hamiltonian with singlet-bond and triplet-current interactions. 
      The blue squares are the phase boundary of the triplet $d_{x^2-y^2}$ density wave order. The red circles denote the phase boundary of the singlet $p_x$ density wave order. The black-hatched region represents the mixed-ordered area where two orders coexist.
      The singlet-bond and triplet-current interaction strengths are denoted by $g_1$ and $g_2$, respectively.
      The left and right insets show respectively the triplet $d_{x^2-y^2}$ and singlet $p_x$ density wave orders on the square lattice.
  }\label{fig:fig1}
\end{figure}

At the mean-field level, the singlet $p_x$ and triplet $d_{x^2-y^2}$ density waves are favored by the singlet-bond and triplet-current interactions, respectively.
A recent Majorana QMC study of the half-filled SU($N$) fermions with singlet-bond interactions on the honeycomb lattice demonstrated a quantum phase transition from the Dirac semimetal to the spin dimerized insulator as the interaction is increased \cite{li2017fermion}.
For comparison, a SU($N$) fermion model with triplet-current interactions was studied by using the projector determinant QMC (PQMC) method, where the doping and values of $N$ can strongly affect the triplet $d_{x^2-y^2}$ density wave order of the ground state \cite{capponi2007spin}. 
However, the model with both the singlet-bond and triplet-current interactions receives much less attention.
A systematic nonperturbative study of its ground state properties is still missing. 
In particular, it is not clear how the two interaction terms compete and induce the quantum phase transition between the singlet $p_x$ and triplet $d_{x^2-y^2}$ density waves.
In this paper, we propose to study the SU($N$) generalization of a spin-$\frac{1}{2}$ model \cite{wu2005sufficient} that includes both the singlet-bond and triplet-current interactions.
We shall conduct a sign-problem-free PQMC study of the half-filled $N=4$ model on the square lattice. 
The zero-temperature phase diagram, Fig.~\ref{fig:fig1}, is obtained as a function of the singlet-bond and triplet-current interaction strengths.
In the weak triplet-current interaction regime, the triplet $d_{x^2-y^2}$ density wave order is observed.
It is shown that the increase of the triplet-current interaction eventually drives the system into an insulating singlet $p_x$ density wave state.
This transition is accompanied by an intermediate state where the two orders coexist. 
Furthermore, the single-particle gap and spectrum are investigated by the unequal-time Green's function and analytical continuation methods.

The rest of this paper is organized as follows. 
In Sec.~\ref{sec:sec2}, we introduce the SU($N$)-symmetric Hamiltonian with singlet-bond and triplet-current interactions, and briefly review the scheme of PQMC simulations. 
The phase diagram of the half-filled $N=4$ model is discussed in Sec.~\ref{sec:sec3}. Subsequently in Sec.~\ref{sec:sec4}, the single-particle gap and spectrum are studied. 
The conclusions are drawn in Sec.~\ref{sec:sec5}.

\section{Model and method}\label{sec:sec2}
Spin-$\frac{1}{2}$ fermions on the lattice bond can construct either the singlets or the triplets. 
Thus, the spin-$\frac{1}{2}$ model Hamiltonian of the singlet-bond and triplet-current interactions is defined as \cite{wu2005sufficient}
\begin{equation}
    \label{eq:h}
    \begin{aligned} 
        H_I = \sum_{\langle{ij}\rangle}-\frac{g_{1}}{2}(c_{i}^{\dagger}\sigma_0c_{j}+\mathrm{H.c.})^2-\frac{g_{2}}{2}(ic_{i}^{\dagger}\frac{\vec{\sigma}}{2}c_{j}+\mathrm{H.c.})^2,
    \end{aligned}
\end{equation}
where $\langle{ij}\rangle$ represents the nearest-neighbor sites and $c_i^{\dagger}=(c_{i\uparrow}^{\dagger},c_{i\downarrow}^{\dagger})$ is the fermion creation operator at site $i$ on the square lattice.
$\sigma_0$ represents a $2\times2$ identity matrix, and $\vec{\sigma}=(\sigma_x,\sigma_y,\sigma_z)$ where $\sigma_x$, $\sigma_y$ and $\sigma_z$ are the Pauli matrices. 
One might argue that the $g_{1}$ term favors the singlet $p$-wave density wave order, while the $g_2$ term favors the triplet $d$-wave density wave order in the mean-field theory. 
However, previous QMC studies \cite{capponi2007spin,li2017fermion} have shown that 
the ground state
can be the antiferromagnetic (AFM) order or the superconducting (SC) order
in the half-filled spin-$\frac{1}{2}$ model with only $g_1$ or $g_2$ terms.
In fact, the interaction Hamiltonian~\eqref{eq:h} can be rewritten as the sum of the pair hopping, density-density, and Heisenberg exchange interactions,
\begin{equation}
    \label{eq:h2}
    \begin{aligned}
        H_I = \sum_{\langle{ij}\rangle}&-(g_1+\frac{3}{4}g_2)(c_{i\uparrow}^{\dagger}c_{i\downarrow}^{\dagger}c_{j\downarrow}c_{j\uparrow}+\mathrm{H.c.}) \\
        &+(\frac{1}{2}g_1+\frac{3}{8}g_2)(n_i-1)(n_j-1) \\
        &+(2g_1-\frac{1}{2}g_2)\vec{S}(i)\cdot\vec{S}(j),
    \end{aligned}
\end{equation}
where $n_i=c_{i\uparrow}^{\dagger}c_{i\uparrow}+c_{i\downarrow}^{\dagger}c_{i\downarrow}$ and $\vec{S}(i)=c_{i}^{\dagger}\frac{\vec{\sigma}}{2}c_{i}$ are the fermion number operator and spin operator at site $i$, respectively. 
In particular, the three terms on the right hand side of Eq.~\eqref{eq:h2} favor the superconducting state, charge density wave and spin density wave, respectively. 

Consider the SU($N$) generalization with spinors of $2N$ components $c_{i,\alpha}^{\dagger}=(c_{i\uparrow,\alpha}^{\dagger},c_{i\downarrow,\alpha}^{\dagger})$, $\alpha=1,2\dots N$, replacing $c_i^{\dagger}=(c_{i\uparrow}^{\dagger},c_{i\downarrow}^{\dagger})$ in the spin-$\frac{1}{2}$ model.
We obtain the generalized SU($N$)-symmetric singlet bond operator
\begin{equation}
    M_{ij} = \frac{1}{\sqrt{N}}\sum_{\alpha=1}^{N}(c_{i,\alpha}^{\dagger}\sigma_0c_{j,\alpha}+\mathrm{H.c.}),
\end{equation}
and triplet current operator
\begin{equation}
    \vec{N}_{ij} = \frac{i}{\sqrt{N}}\sum_{\alpha=1}^{N}(c_{i,\alpha}^{\dagger}\frac{\vec{\sigma}}{2}c_{j,\alpha}-\mathrm{H.c.}).
\end{equation}
So, the SU($N$)-symmetric Hamiltonian with singlet-bond and triplet-current interactions is defined as
\begin{equation}\label{eq:hN}
    H = -t\sum_{\langle{ij}\rangle,\alpha}(c_{i,\alpha}^{\dagger}c_{j,\alpha} +\mathrm{H.c.}) + \sum_{\langle{ij}\rangle}[-\frac{g_{1}}{2}M_{ij}^2-\frac{g_{2}}{2}\vec{N}_{ij}^2].
\end{equation}
At large-$N$ limit, the Hubbard-Stratonovich fields, $\chi_{ij}$ and $\vec{J}_{ij}$, defined on every lattice bond 
can factorize the terms $M_{ij}^2$ and $\vec{N}_{ij}^2$, corresponding to the mean-field order parameters $\chi$ and $\vec{J}$ (see Appendix~\ref{sec:supp:mean}).
For the triplet $d_{x^2-y^2}$ density wave order
we can derive the mean-field dispersion relation 
$E_k(\vec{J})=\pm\sqrt{\epsilon_k^2+(\vec{\Delta}/2)^2}$ 
with $\vec{\Delta}=-(\cos{k_x}-\cos{k_y})\vec{J}/4$;
hence there exist $4N$ low-energy anisotropic Dirac cones located at ($\pi/2,\pm\pi/2$) when taking into account the spin degeneracy.
In particular, the dispersion relation is a linear function of $\vec{\kappa}$ around the Dirac point $\vec{K}=(\pi/2,\pi/2)$
where $\vec{\kappa}$ is the deviation from $\vec{K}$. 
Conversely, the mean-field dispersion relation of the singlet $p_x$ density wave opens an energy gap at all wavevectors.
Using the mean-field ansatz of the singlet $p_x$- and triplet $d_{x^2-y^2}$-wave orderings, we can solve the saddle-point equations self-consistently.
Certainly, the singlet $p_x$ and triplet $d_{x^2-y^2}$ density wave states emerge when $g_1>0,g_2=0$ and $g_1=0,g_2>0$, respectively. 
Nevertheless, by increasing $g_1$ at $g_2>0$, the singlet $p_x$-wave order is formed at a nonzero $g_1$.
As $g_1$ is further increased, the $d_{x^2-y^2}$ density wave order is gradually suppressed and the two density wave orders coexist. 
The problem of coexistence associated with the coupling of order parameters can be proved by using a phenomenological Ginzburg-Landau (GL) description,
as shown in Appendix~\ref{sec:supp:mean}.

Below, let us briefly describe the PQMC method in the determinant formalism \cite{blankenbecler1981monte,hirsh1985two,Assaad2008Worldline}.
The model Hamiltonian \eqref{eq:hN} can be simulated without a sign problem by using the Kramer's time-reversal invariant decomposition \cite{wu2005sufficient},
\begin{equation}\label{eq:hs-decomp}
    \begin{aligned}
        e^{\Delta\tau gM_{ij}^2} &= \sum_{l,s=\pm1}\frac{\gamma_l}{4}e^{s\eta_l\sqrt{\Delta\tau g}M_{ij}}+\mathcal{O}(\Delta\tau^4), \\
        e^{\Delta\tau g\vec{N}_{ij}^2} &= \prod_{a=x,y,z}\sum_{l_a,s_a=\pm1}\frac{\gamma_{l_a}}{4}e^{s_a\eta_{l_a}\sqrt{\Delta\tau g}{N}_{ij,a}}+\mathcal{O}(\Delta\tau^4),
    \end{aligned}
\end{equation}
where $\gamma_l = 1+\frac{\sqrt{6}}{3}l$ and $\eta_l=\sqrt{2(3-\sqrt{6}l)}$  \cite{wu2005sufficient,Assaad2008Worldline}. 
In this case, the discrete auxiliary fields have $4^4=256$ possible choices on every bond. 
Moreover, Eq.~\eqref{eq:hs-decomp} allows us to decouple the fermion operators $c_{i,\alpha}$ and $c_{i,\beta\ne\alpha}$, corresponding to different subspaces of the Hilbert space. 
Therefore, the propagation operator is rewritten as
\begin{equation}
    \begin{aligned}
        \langle{\Psi_T}|{e^{M\Delta\tau H}}|{\Psi_T}\rangle = \sum_{\{l,s\}}\gamma(\{l\})\Big[\det P^{\dagger}\prod_{p=1}^{M} B_{p} P \Big]^{N},
    \end{aligned}
\end{equation}
where $\gamma(\{l\})=\prod_{p=1}^{M}\frac{\gamma_{l_p}}{4}$ and $B_{p}=e^{-\Delta\tau H_{0,\alpha}} e^{-\Delta\tau H_{I,\alpha}}$ is defined in the subspace of flavor $\alpha$. 
The rectangular matrix, $P$, characterizes the Slater determinant of the trial wave function $|\Psi_T\rangle$.
More implementation details of the algorithm can be found in Refs.~\cite{Assaad2008Worldline,wang2014competing} and in the source codes \cite{SourceCode}.
Our PQMC simulations are performed on 24 CPU cores with 500 Monte Carlo steps for warming up and 500 steps for measurements on each core (see details in Appendix~\ref{sec:supp:qmc}). The square lattice is subject to the periodic boundary condition. 
The Trotter decomposition step $\Delta\tau=0.1$ and projection time $M\Delta\tau=24$ are used. 
The measurements of physical observables are performed close to $M\Delta\tau/2$ after projecting onto the ground state.

\begin{figure}
  \centering
  \includegraphics[width=0.9\linewidth]{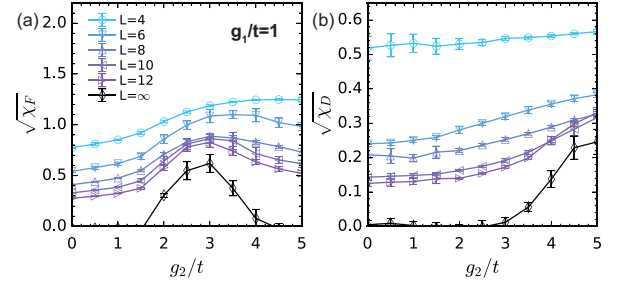}
  \caption{
      Order parameters as a function of $g_2$ at $g_1/t=1$. (a) Triplet $d_{x^2-y^2}$ and (b) singlet $p_x$ density wave order parameters. 
      Black curves represent the extrapolated order parameters in the $1/L\to0$ limit.
  }\label{fig:fig2}
\end{figure}

\section{Phase diagram}\label{sec:sec3}

Generally, the spin current operators are denoted by $\vec{J}_{i,\hat{e}_a} \equiv \vec{N}_{i,i+\hat{e}_a}$ where $a=x,y$ and $\hat{e}_{x},\hat{e}_{y}$ represent the primitive lattice vectors of the square lattice. 
The structure factor of the triplet $d_{x^2-y^2}$ density wave state is then defined as
\begin{equation}
    \chi_{F} = \sum_{a,b=x,y}\chi_{F,ab} = \sum_{a,b=x,y}\frac{\varepsilon_{ab}}{L^4}\sum_{ij}\langle{J_{i,\hat{e}_a}J_{j,\hat{e}_b}}\rangle e^{i\vec{Q}\cdot\vec{r}},
\end{equation}
where $\vec{Q}=(\pi,\pi)$, $\vec{r}=\vec{r}_i-\vec{r}_j$, $\varepsilon_{xx}=\varepsilon_{yy}=1$ and $\varepsilon_{xy}=\varepsilon_{yx}=-1$.
As for the singlet $p_x$ density wave order, the kinetic bond operators are expressed as $d_{i,\hat{e}_a}\equiv M_{i,i+\hat{e}_a}$. 
Consider the spin dimerization along the $\hat{e}_{x}$ and $\hat{e}_{y}$ directions.
We define the structure factor of the singlet $p_x$ density wave state as
\begin{equation}
    \chi_{D} = \sum_{a=x,y}\chi_{D,a} = \sum_{a=x,y}\frac{1}{L^4}\sum_{ij}\langle{d_{i,\hat{e}_a}d_{j,\hat{e}_a}}\rangle e^{i\vec{Q}_a\cdot\vec{r}},
\end{equation}
where $\vec{Q}_x=(\pi,0)$ and $\vec{Q}_y=(0,\pi)$.

Considering the SC and AFM instabilities described by Eq.~\eqref{eq:h2}, we also measure the structure factors of the SC order,
\begin{equation}
    \chi_{C} = \frac{1}{L^4}\sum_{ij}\sum_{\alpha\beta}\langle{c_{i\uparrow,\alpha}^{\dagger}c_{i\downarrow,\alpha}^{\dagger}c_{j\downarrow,\beta}c_{j\uparrow,\beta}}\rangle,
\end{equation}
and the AFM order,
\begin{equation}
    \chi_{S} = \frac{1}{L^4}\sum_{ij}\langle{\vec{S}(i)\cdot\vec{S}(j)}\rangle e^{i\vec{Q}\cdot\vec{r}},
\end{equation}
where $\vec{S}(i)=\sum_{\alpha}c_{i,\alpha}^{\dagger}\frac{\vec{\sigma}}{2}c_{i,\alpha}$.

For the purpose of simplicity, we plot the order parameters as a function of $g_2$ while fixing $g_1/t=1$. 
As shown in Fig.~\ref{fig:fig2}(a), following the successive increase of $g_2$, the triplet $d_{x^2-y^2}$ density wave order parameter, $\sqrt{\chi_F}$, increases at first and then decreases for lattice sizes $L\geqslant 6$. 
Extrapolation to the limit of $1/L\to0$ shows that the triplet $d_{x^2-y^2}$ density wave order starts to appear at around $g_2/t\approx 1.5$. 
As further increasing $g_2$, the order parameter in the $1/L\to0$ limit becomes nonmonotonic: 
it keeps increasing until it reaches the maximum around $g_2/t\approx 3$. 
After that, it declines steadily to zero when $g_2/t\gtrsim 4$. 
Meanwhile, the analysis for the singlet $p_x$ density wave order parameter can be carried out in parallel. 
As shown in Fig.~\ref{fig:fig2}(b), the singlet $p_x$ density wave order develops when $g_2/t\gtrsim 3$, which is beyond the mean-field theory.
Note that near $g_2=0$ the extrapolated values of $\sqrt{\chi_D}$ are very small. 
It is difficult to judge whether the $p_x$-wave order vanishes.
Nevertheless, later in Sec.~\ref{sec:sec4}, the single-particle gap data show nonzero values, and thus are consistent with weak $p_x$-wave orderings.

In the large-$g_2$ regime, the vanishing triplet $d_{x^2-y^2}$ and nonzero singlet $p_x$ density wave order parameters are somewhat counterintuitive, 
because the $g_2$ term in the Hamiltonian~\eqref{eq:hN} favors the triplet $d_{x^2-y^2}$ density wave order at the mean-field level. 
To explain why increasing $g_2$ favors the $p_x$ density wave order and suppresses the $d_{x^2-y^2}$ density wave order, let us discuss an intuitive picture as follows. 
Denote the single occupancy, double occupancy and empty states by $|\cdot\rangle$, $|\!\cdot\!\cdot\rangle$ and $|\empty\rangle$, respectively. 
The current state of a two-site system is essentially the superposition $\ketdz + e^{i\phi}\ketoo + e^{2i\phi}\ketzd$ with a phase difference $\phi$, while the bond state is the superposition $\ketdz + \ketoo + \ketzd$ without the phase difference. 
The key argument is that at large $g_2$ the virtual hopping process brought by the kinetic term $H_{0,\alpha}$ does not cause any phase difference and thus favors the bond state. 
However, each $H_{0,\alpha}$ only acts on the subspace of flavor $\alpha$, which is factor-$N$ times smaller than the interaction terms in the SU($N$)-symmetric Hamiltonian~\eqref{eq:hN}. 
Thus the kinetic energy gain for the bond state is neglectable for $1/N\to0$. 
Hence, from the energy perspective, the singlet-bond state is the ground state when both conditions of large $g_2$ and small $N$ are met. 
Also, it is worthwhile to be reminded that $N$ cannot be arbitrarily small like $N=1$.

\begin{figure}
  \centering
  \includegraphics[width=0.96\linewidth]{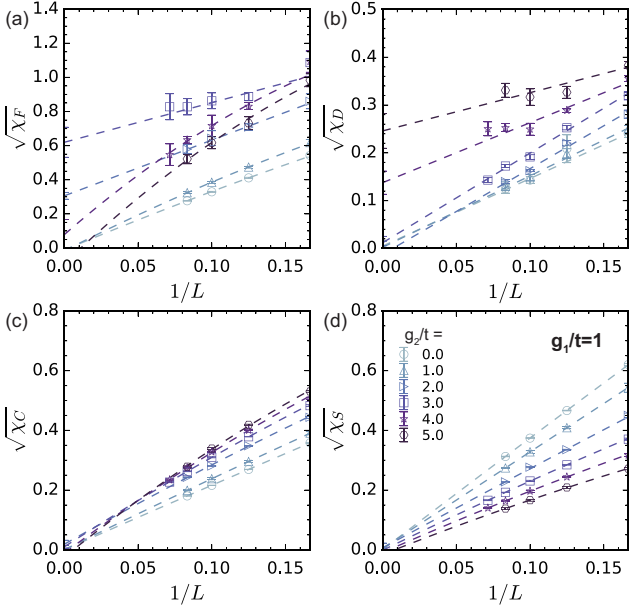}
  \caption{
      The finite-size extrapolation of order parameters as $g_2$ varies and $g_1/t=1$. (a) Triplet $d_{x^2-y^2}$ and (b) singlet $p_x$ density wave order parameters. (c) SC and (d) AFM order parameters.
  }\label{fig:fig3}
\end{figure}

Moreover, according to Fig.~\ref{fig:fig2}, the singlet $p_x$ and triplet $d_{x^2-y^2}$ density wave order parameters in the $1/L\to0$ limit are both nonzero for $3.0\lesssim g_2/t \lesssim 4.0$. 
In other words, the quantum phase transition between the singlet $p_x$ and triplet $d_{x^2-y^2}$ density wave states has an intermediate region of coexistence as tuning $g_2$.
The coexistence region of two orders is usually termed as the mixed-ordered area or coexisting phase in the literatures \cite{watanabe1985phase,anisimov1981phase}.
As discussed in Appendix~\ref{sec:supp:mean}, coexistence of the singlet $p_x$ and triplet $d_{x^2-y^2}$ density wave order parameters is allowed in the GL theory.

Details of the finite-size extrapolation are shown in Fig.~\ref{fig:fig3}, 
where the polynomial curve fitting of $1/L$ is employed.
In the presence of long-range correlations defined on the lattice bond, 
the order parameters of $L=4n$ and $L=4n+2$ have vastly different values due to strong finite-size effects. 
In this case, we fit our data to the linear function of $1/L$ to average the finite-size effects, as shown in Figs.~\ref{fig:fig3}(a) and \ref{fig:fig3}(b).
In contrast, $\sqrt{\chi_C}$ and $\sqrt{\chi_S}$ are well described by the polynomials of $1/L$, and their extrapolated results prove the absence of SC and AFM orders, as Figs.~\ref{fig:fig3}(c) and (d) show. 

Next, we consider the correlation ratio which concerns the ratio between structure factors at an
ordering wavevector and its nearest wavevector. For example, the correlation ratio of the triplet $d_{x^2-y^2}$ density wave is defined as
\begin{equation}
    R_{F}(L) = 1 - \frac{1}{4}\sum_{a=x,y}\left(\frac{\chi_F(\vec{Q}-d\vec{q}_a)}{\chi_F(\vec{Q})}+\frac{\chi_F(\vec{Q}+d\vec{q}_a)}{\chi_F(\vec{Q})}\right),
\end{equation}
where $d\vec{q}_x=(2\pi/L,0)$ and $d\vec{q}_y=(0,2\pi/L)$. 
Similarly, the correlation ratio of the singlet $p_x$ density wave is defined as
\begin{equation}
    R_{D}(L) = 1 - \frac{1}{4}\sum_{a,b=x,y}\frac{\chi_D(\vec{Q}_a+d\vec{q}_b)}{\chi_D(\vec{Q}_a)}.
\end{equation}
In the ordered phase, the correlation ratio goes to one in the $1/L\to0$ limit; whereas in the disordered phase, it goes to zero. 

\begin{figure}
  \centering
  \includegraphics[width=0.96\linewidth]{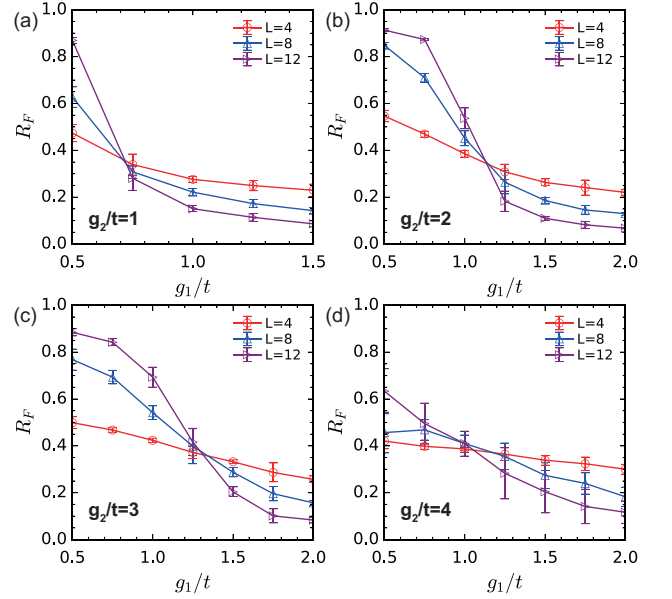}
  \caption{
    The correlation ratio of the triplet $d_{x^2-y^2}$ density wave as a function of $g_1$ when $g_2$ is fixed. (a) $g_2/t=1$, (b) $g_2/t=2$, (c) $g_2/t=3$ and (d) $g_2/t=4$.
  }\label{fig:fig4}
\end{figure}

Figure~\ref{fig:fig4} shows $R_F$ as a function of $g_1$ for some fixed values of $g_2$. 
Here, we fit the data to polynomial functions of $g_1$, and estimate the crossing point between the fitted curves of $R_F(L)$ and $R_F(L+4)$ by using the bootstrap method.
Since the correlation ratio is a renormalization-group-invariant quantity, we view the average crossing points as the phase transition points regardless of scaling corrections \cite{Parisen2015Fermionic}.
Overall, Fig.~\ref{fig:fig4} indicates that the critical points of the triplet $d_{x^2-y^2}$ density wave order are $g_1/t=0.71(2),1.14(2),1.31(4),1.06(15)$ for the fixed $g_2/t=1,2,3,4$ respectively. 
Similar analyses are carried out in other parameter regimes, and the results are plotted in blue squares, as shown in Fig.~\ref{fig:fig1}.
In our simulations, however, the $R_D$ data exhibit large error bars; 
as a consequence, we could not obtain the crossing points of the $R_D$ curves. 
Alternatively, we can extract the critical point $g_{c}$ by fitting the extrapolated data to 
$\sqrt{\chi_D}\sim(g-g_{c})^\beta$, as shown in Appendix~\ref{sec:supp:data}.
Critical points of the singlet $p_x$-wave order are denoted by the red circles in Fig.~\ref{fig:fig1}.
In the phase diagram, as tuning $g_1$, the discrepancy in between the blue squares and red circles indicates a disordered ground state with vanishingly small order parameters, which is attributed to a tie between the singlet-bond and triplet-current interactions without either side winning in this parameter regime.

\begin{figure}
    \includegraphics[width=0.96\linewidth]{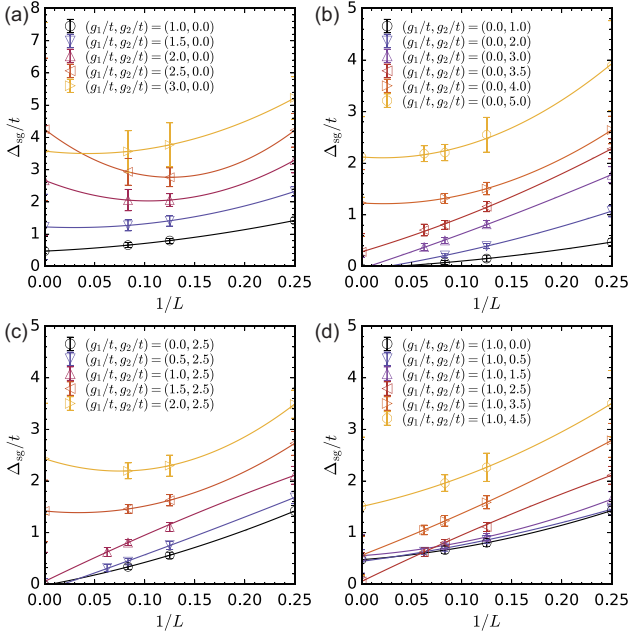}
    \caption{
        The finite-size extrapolation of the single-particle gap $\Delta_\text{sg}$ for various parameters $(g_1,g_2)$:
        (a) varying $g_1/t$ and fixing $g_2/t=0$; (b) fixing $g_1/t=0$ and varying $g_2/t$; 
        (c) varying $g_1/t$ and fixing $g_2/t=2.5$; (d) fixing $g_1/t=1$ and varying $g_2/t$.
        The quadratic polynomial fitting is applied.
    }\label{fig:fig5}
\end{figure}

\section{Gap opening mechanism}\label{sec:sec4}
So far we have analyzed the equal-time observables, which gave us the phase diagram 
including two kinds of nonzero angular momentum density wave phases and a mixed-ordered area.
In this section, we investigate the single-particle gap $\Delta_\text{sg}$ and spectrum $A(\vec{k},\omega)$
so as to further clarify the density wave phases.

Physics of the singlet $p_x$ and triplet $d_{x^2-y^2}$ density wave states is very different according to the mean-field analysis in Appendix~\ref{sec:supp:mean}:
The triplet $d_{x^2-y^2}$ density wave state possesses Dirac fermion spectrum, which is gapless at the wavevector $\vec{K}$;
whereas the singlet $p_x$ density wave state opens a gap at all wavevectors.

We consider the unequal-time Green's function as
\begin{equation}
    G(\vec{k};\tau) = \frac{1}{L^2}\sum_{ij}\langle c_{i}(\tau)c_{j}^{\dagger}(0) \rangle e^{i\vec{k}\cdot\vec{r}},
\end{equation}
and extract $\Delta_\text{sg}$ of momentum $\vec{k}$ using $G(\vec{k};\tau)\sim e^{-\Delta_\text{sg}\tau}$.
Since the minimal single-particle gap is located at the Dirac point $\vec{K}$, we fit $\ln G(\vec{K};\tau)$ of the $4n\times 4n$ square lattice to a linear function of $\tau$ within the range where the data of $\ln{G}$ versus $\tau$ show asymptotic linear behavior. 
Then the finite-size extrapolation of $\Delta_\text{sg}$ is performed using the quadratic polynomial functions of $1/L$. 

In Figs.~\ref{fig:fig5}(a) and \ref{fig:fig5}(b), extrapolations of $\Delta_\text{sg}$ along the $g_1$ axis and $g_2$ axis are plotted, respectively.
For the $g_1$ axis, QMC results always give nonzero extrapolated values of $\Delta_\text{sg}$, which indicates the singlet $p_x$ density wave orderings at small $g_1$.
In contrast, for the $g_2$ axis the extrapolated $\Delta_\text{sg}$ is equal to zero when $g_2/t\leqslant 3$, 
which is consistent with the triplet $d_{x^2-y^2}$ density wave order.
After that, the system enters the mixed-ordered area and $\Delta_\text{sg}>0$ for $g_2/t> 3$.
Theoretically, the singlet $p_x$-wave ordering breaks the nodal point's energy degeneracy of the $d_{x^2-y^2}$-wave order; and thus the mixed-ordered area is gapped at $\vec{K}$.

Figures~\ref{fig:fig5}(c) and \ref{fig:fig5}(d) show the extrapolation of $\Delta_\text{sg}$ for nonzero $g_1$ and $g_2$.
When $g_2/t=2.5$ and $g_1$ increases, there is a transition from the triplet $d_{x^2-y^2}$ density wave to the singlet $p_x$ density wave in the phase diagram. 
In this case, we find $\Delta_\text{sg}=0$ for $g_1/t\leqslant 1$ and $\Delta_\text{sg}>0$ for $g_1/t\geqslant 1.5$, as shown in Fig.~\ref{fig:fig5}(c).
For comparison, in Fig.~\ref{fig:fig5}(d), when $g_1/t=1$ and $g_2$ increases, $\Delta_\text{sg}$ is nonzero for small $g_2$, but $\Delta_\text{sg}$ drops to zero at $g_2/t=2.5$, corresponding to the transition from the singlet $p_x$ density wave to the triplet $d_{x^2-y^2}$ density wave.
Further increasing $g_2$ reopens the energy gap at $g_2/t= 3.5$, meaning that the system enters the mixed-ordered area and eventually reenters the pure singlet $p_x$ density wave phase.
These results are consistent with the phase boundary of the triplet $d_{x^2-y^2}$ density wave order.

\begin{figure}
    \includegraphics[width=0.96\linewidth]{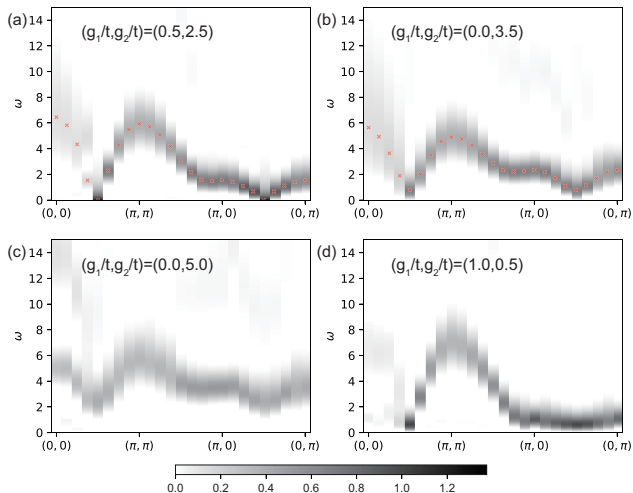}
    \caption{
        The single-particle spectral function $A(\vec{k},\omega)$ on a $16\times16$ square lattice
        along a path in the reciprocal lattice. 
        (a) $(g_1/t,g_2/t)=(0.5,2.5)$ is in the pure triplet $d_{x^2-y^2}$ density wave phase.
        (b) $(g_1/t,g_2/t)=(0,3.5)$ is in the mixed-ordered area.
        The red-product sign represents the position of the maximum value, $\arg \max_{\omega} A(\omega)$.
        (c) $(g_1/t,g_2/t)=(0,5)$ is in the pure singlet $p_x$ density wave phase.
        (d) $(g_1/t,g_2/t)=(1,0.5)$ is of a weak $p_x$-wave ordering.
        We have normalized $A(\vec{k},\omega)$ of each $\vec{k}$ to unity.
    }\label{fig:fig6}
\end{figure}

Previous studies have presented the single-particle spectrum $A(\vec{k},\omega)$ to confirm the semimetal character of the $d_{x^2-y^2}$-wave order \cite{assaad2005phase,capponi2007spin}.
However, $A(\vec{k},\omega)$ in the mixed-ordered area has not been investigated.
In our simulations, we perform the analytical continuation that utilises sparse modeling approach \cite{otsuki2017sparse} to derive $A(\vec{k},\omega)$ from the equation
\begin{equation}
    G(\vec{k},\tau) = \int_{-\infty}^{+\infty}d\omega\,\theta(\omega)e^{-\tau\omega}A(\vec{k},\omega),
\end{equation}
where $\theta(\omega)$ is the step function.

Before presenting numerical results, let us show the anisotropic Dirac cone in the spectrum of the triplet $d_{x^2-y^2}$ density wave order. 
Expanding the mean-field Hamiltonian (see Appendix~\ref{sec:supp:mean}) at $\vec{K}$ as a function of $\vec{\kappa}$,
we obtain $H=2\sqrt{2}t\sigma_0\kappa_{\perp}\tau_z+\frac{\sqrt{2}}{8}(\vec{J}\cdot\vec{\sigma}) \kappa_{\parallel}\tau_y$. 
Here $\hat\kappa_{\perp}=(\hat{\kappa}_x+\hat{\kappa}_y)/\sqrt{2}$, $\hat\kappa_{\parallel}=(\hat{\kappa}_x-\hat{\kappa}_y)/\sqrt{2}$, $\vec{J}$ is the mean-field order parameter, and $\tau_z,\tau_y$ are the Pauli matrices defined in the $(c_{k},c_{k+Q})$ basis.
Thus, we arrive at two different velocities $v_{\perp}=2\sqrt{2}t$ and $v_{\parallel}=\frac{\sqrt{2}}{8}J$ which characterize the anisotropy of Dirac cone. 
In addition, the ratio $v_{\parallel}/v_{\perp}$ gives the mean-field order parameter \cite{capponi2007spin}.

As shown in Fig.~\ref{fig:fig6}(a), $A(\vec{k},\omega)$ near the Dirac point $\vec{K}=(\pi/2,\pi/2)$ clearly shows the anisotropic Dirac cone and gapless single-particle excitations.
Fitting the position of the maximum value, $\arg \max_{\omega} A(\omega)$, to a linear function of $\vec{\kappa}$, we obtain the ratio $v_{\parallel}/v_{\perp}\approx 0.27$.
For comparison, Fig.~\ref{fig:fig6}(b) shows $A(\vec{k},\omega)$ in the mixed-ordered area, which has several features. 
For instance, an energy gap opens at $\vec{K}$, which is consistent with the extrapolation of $\Delta_\text{sg}$. 
Remarkably, the velocities around $\vec{K}$ remain anisotropic and the ratio is $v_{\parallel}/v_{\perp}\approx 0.32$.
In contrast, inside the pure singlet $p_x$ density wave phase, 
the energy gap at all wavevectors is evident, and $\arg\max_{\omega} A(\omega)$ is a quadratic function of $\vec{\kappa}$ around $\vec{K}$, as Fig.~\ref{fig:fig6}($c$) shows.
Furthermore, the data in Fig.~\ref{fig:fig6}(d) are significantly different from the data in Fig.~\ref{fig:fig6}($c$).
In particular, $\arg\max_{\omega} A(\omega)$ along the $\hat{\kappa}_{\parallel}$ direction around $\vec{K}$ is very flat and the energy gap at $\vec{K}$ is very small, 
which shows the tendency towards the Fermi surface of noninteracting limit and reflects the weak $p_x$-wave ordering. 
Therefore, Fig.~\ref{fig:fig6}(d) shows $A(\vec{k},\omega)$ of a weak singlet $p_x$ density wave order.

\section{Conclusions}\label{sec:sec5}
In summary, we have performed the PQMC simulations of a SU($N$)-symmetric Hamiltonian with singlet-bond and triplet-current interactions on the square lattice. 
We find the gapped singlet $p_x$ and gapless triplet $d_{x^2-y^2}$ density wave states in the half-filled $N=4$ model.
Without the singlet-bond interaction, the mean-field ground state is the triplet $d_{x^2-y^2}$ density wave order for any nonzero triplet-current interaction strengths.
In contrast, our QMC simulations show a transition to the singlet $p_x$ density wave when the triplet-current interaction strength is increased, which is beyond the mean-field theory. 
This transition is accompanied by a gapped mixed-ordered area where two orders coexist, and the coexistence of two competing orders is explained in the GL description.
After turning on the singlet-bond interaction, there is a transition from the triplet $d_{x^2-y^2}$ to the singlet $p_x$ density wave phases. 
In this case, however, the ground state is disordered in between the two ordered phases.
Furthermore, we investigate the single-particle spectrum by employing the recently developed sparse modeling approach. 
For the triplet $d_{x^2-y^2}$ density wave, the anisotropic Dirac cone is observed in the spectrum.
On the other hand, the spectrum of the singlet $p_x$ density wave shows a parabolic shape around the Dirac point and has the energy gap at all wavevectors.
As for the mixed-ordered area, an energy gap is opened and the velocities remain anisotropic at the Dirac point.

\section*{Acknowledgments}
This work is financially supported by
the National Natural Science
Foundation of China under Grants No. 11874292,
No. 11729402, and No. 11574238.
We acknowledge the support of the Supercomputing Center of Wuhan University.
C.W. is
supported by the National Natural Science Foundation of China under the Grants No. 12174317
and No. 12234016.

\appendix

\appsection{Supplementary data}\label{sec:supp:data}

\begin{figure*}
    \includegraphics[width=0.9\linewidth]{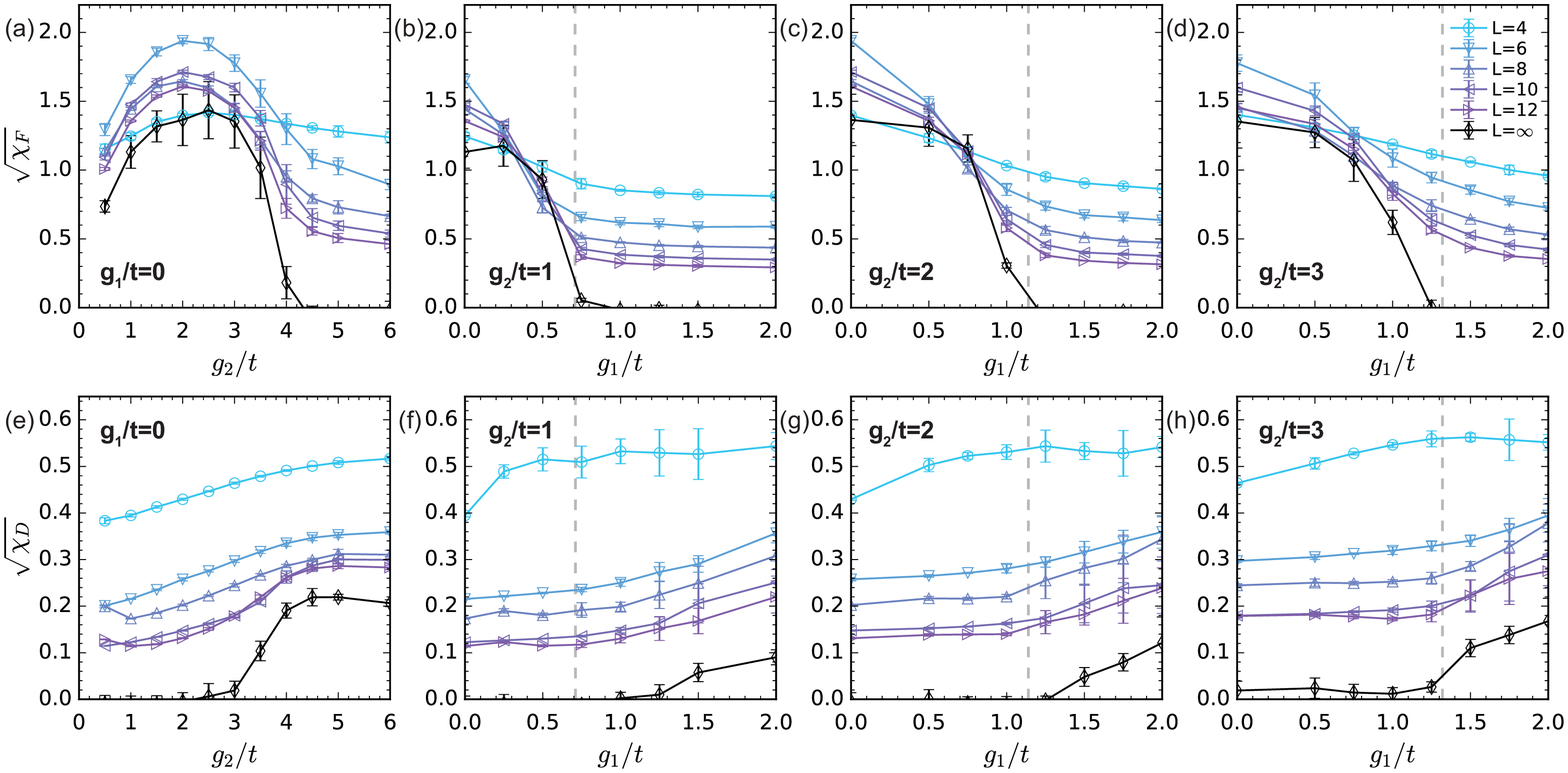}
    \caption{Order parameters as a function of $g_1$ ($g_2$) when $g_2$ ($g_1$) is fixed. [(a)-(d)] Triplet $d_{x^2-y^2}$ and [(e)-(h)] singlet $p_x$ density wave order parameters.
    Dashed vertical lines represent the phase transition points extracted from the correlation ratio $R_F$.
    }\label{fig:supp:fig7}
\end{figure*}

Without the singlet-bond interaction, i.e., at $g_1=0$,
$\sqrt{\chi_F}$ and $\sqrt{\chi_D}$ as a function of $g_2$ are plotted in Figs.~\ref{fig:supp:fig7}(a) and \ref{fig:supp:fig7}(e), respectively.
Here $\sqrt{\chi_F}$ and $\sqrt{\chi_D}$ represent the order parameters of the triplet $d_{x^2-y^2}$ and singlet $p_x$ density waves, respectively.
In Fig.~\ref{fig:supp:fig7}(a), the extrapolated value of $\sqrt{\chi_F}$ increases until it reaches the maximum at around $g_2/t\approx 2.5$. After that, it drops to zero when $g_2/t\gtrsim 4.5$. 
By contrast, the extrapolated $\sqrt{\chi_D}$ becomes greater than zero when $g_2\gtrsim 3.5$, as Fig.~\ref{fig:supp:fig7}(e) shows. 
By fitting the data to $\sqrt{\chi_D}\sim(g_2-g_{2,c})^\beta$, we obtain the critical point $g_{2,c}=2.7\pm0.16$ of the singlet $p_x$ density wave phase.

For the rest of the data in Fig.~\ref{fig:supp:fig7}, 
we plot the order parameters as a function of $g_1$ while fixing $g_2$.
Additionally, we denote by dashed vertical lines the phase transition points of the triplet $d_{x^2-y^2}$-wave order.
At $g_2/t=1$, the data of $\sqrt{\chi_F}$ and $\sqrt{\chi_D}$ are plotted in Figs.~\ref{fig:supp:fig7}(b) and \ref{fig:supp:fig7}(f), respectively.
In this case, $\sqrt{\chi_F}$ goes to zero when $g_1/t\gtrsim0.75$,
whereas nonzero values of $\sqrt{\chi_D}$ only appear after $g_1/t\gtrsim1.25$.
Therefore, both order parameters are vanishingly small and the ground state is disordered in the parameter regime around $g_1/t\approx1$.
Moreover, in this regime the single-particle gap $\Delta_\text{sg}>0$, as discussed in Sec.~\ref{sec:sec4} of the main text.
Similarly, Figs.~\ref{fig:supp:fig7}(c) and \ref{fig:supp:fig7}(g) and Figs.~\ref{fig:supp:fig7}(d) and \ref{fig:supp:fig7}(h) show the data at $g_2/t=2$ and $g_2/t=3$, respectively.
The disordered ground state is also seen at around $g_1/t\approx1.25,g_2/t\approx2$.

\appsection{Details of the mean-field calculation}\label{sec:supp:mean}
We formulate the partition function in a path integral \cite{coleman2015introduction},
\begin{equation}
    \calZ = \int \calD[\bar{c},c]\exp[-\int_0^{\beta}d\tau L],
\end{equation}
where the Lagrangian $L=\sum_{i,\alpha}\bar{c}_{i,\alpha}\partial_{\tau} c_{i,\alpha} + H$ with $\partial_{\tau}c_{i,\alpha}=\partial{c_{i,\alpha}}/\partial{\tau}$ and $H$ given by Eq.~\eqref{eq:hN}. Consider the Hubbard-Stratonovich (HS) transformation that factorizes the fermion interaction terms on every bond.
We rewrite the Lagrangian quadratically \cite{affleck1989largen},
\begin{equation}
    \begin{aligned}
        L \to L + \sum_{\langle ij\rangle} \Big[ \frac{2}{g_1}(\frac{\sqrt{N}\chi_{ij}}{2}-\frac{g_1}{2}M_{ij})^2 \\ 
        + \frac{2}{g_2}(\frac{\sqrt{N}\vec{J}_{ij}}{2}-\frac{g_2}{2}\vec{N}_{ij})^2 \Big].
    \end{aligned}
\end{equation}
So, we obtain the transformed partition function $\calZ = \int \calD[\bar{c},c,\chi,\vec{J}] e^{-S[\bar{c},c,\chi,\vec{J}]}$
where
\begin{equation}
    \begin{aligned}        
    &S[\bar{c},c,\chi,\vec{J}] = \int_0^{\beta}d\tau \sum_{k,\alpha}\bar{c}_{k,\alpha}(\partial_\tau + \epsilon_k)c_{k,\alpha} \\ 
    &+ \sum_{\langle ij\rangle}(\frac{N\chi_{ij}^2}{2g_1}+\frac{N\vec{J}_{ij}^{~2}}{2g_2}-\sqrt{N}\chi_{ij}M_{ij}-\sqrt{N}\vec{J}_{ij}\cdot\vec{N}_{ij}).
    \end{aligned}
\end{equation}
At this point, we can integrate out the fermion fields,
yielding $\calZ = \int \calD[\chi,\vec{J}] e^{-S_E[\chi,\vec{J}]}$.
Here, $S_E$ is the effective action defined as
\begin{equation}\label{eq:effi_expS}
    \begin{aligned}
        &e^{-S_E[\chi,\vec{J}]} = \int \calD[\bar{c},c]e^{-S[\bar{c},c,\chi,\vec{J}]} \\
        &= \det[\partial_\tau + h_E] \exp\Big[-\sum_{\langle ij\rangle}\int_0^{\beta}d\tau\Big(\frac{N\chi_{ij}^2}{2g_1}+\frac{N\vec{J}_{ij}^{~2}}{2g_2}\Big)\Big],
    \end{aligned}
\end{equation}
where we introduce the effective Hamiltonian
\begin{equation}\label{eq:effi_H}
  \begin{aligned}
    h_E=&-t\sum_{\langle ij\rangle,\alpha}(c_{i,\alpha}^{\dagger}c_{j,\alpha} + \mathrm{H.c.}) \\
    &- \sum_{\langle ij\rangle}(\sqrt{N}\chi_{ij}M_{ij}+\sqrt{N}\vec{J}_{ij}\cdot\vec{N}_{ij}).    
  \end{aligned}
\end{equation}
Fourier transform the fields by $c_{j,\alpha}=\frac{1}{\sqrt{N_s}}\sum_{k}c_{k,\alpha}e^{ikR_j}$. 
For the singlet-bond interaction term, we obtain
\begin{equation}\label{eq:sum_mij}
    \begin{aligned}
        \sum_{\langle{ij}\rangle}\sqrt{N}\chi_{ij}M_{ij}&=\sum_{\langle{ij}\rangle,\alpha}(\chi_{ij}c_{i,\alpha}^{\dagger}\sigma_{0}c_{j,\alpha}+\mathrm{H.c.}) \\
        &=\sum_{kk',\alpha}\chi_{k'-k}c_{k',\alpha}^{\dagger}\sigma_{0}c_{k,\alpha},        
    \end{aligned}
\end{equation}
where $\chi_{k'-k}=\frac{1}{2N_s}\sum_{j\delta}\chi_{j,j+\delta} e^{-i(k'-k)R_j}e^{ik\delta}$. 
Similar derivation can be applied to the triplet-current interaction term, and we have
\begin{equation}\label{eq:sum_nij}
    \begin{aligned}
        \sum_{\langle{ij}\rangle}\sqrt{N}J_{ij}N_{ij}&=\sum_{\langle{ij}\rangle,\alpha}(iJ_{ij}c_{i,\alpha}^{\dagger}\frac{\vec{\sigma}}{2} c_{j,\alpha}+\mathrm{H.c.}) \\
        &=\sum_{kk',\alpha}J_{k'-k}c_{k',\alpha}^{\dagger}\frac{\vec{\sigma}}{2} c_{k,\alpha},        
    \end{aligned}
\end{equation}
where $J_{k'-k}=\frac{i}{2N_s}\sum_{j\delta}J_{j,j+\delta} e^{-i(k'-k)R_j}e^{ik\delta}$.
Substituting the Eqs.~\eqref{eq:sum_mij}\eqref{eq:sum_nij} into the effective Hamiltonian \eqref{eq:effi_H} and taking the logarithm of Eq.~\eqref{eq:effi_expS}, we write the effective action in the Matsubara frequencies as
\begin{equation}\label{eq:effi_S}
    \begin{aligned}
        &S_E[\chi,\vec{J}] = N\int_x\Big[\frac{|\chi|^2}{2g_1}+\frac{|\vec{J}|^{2}}{2g_2}\Big] \\
        &-N\mathrm{Tr}\ln[(-i\omega_n+\epsilon_k)\delta_{k,k'}-(\chi_{k'-k}\sigma_0+\vec{J}_{k'-k}\cdot\frac{\vec{\sigma}}{2})]
    \end{aligned}
\end{equation}
where $\int_x=\sum_{\langle{ij}\rangle}\int_0^{\beta}d\tau$, 
and $\epsilon_k=-2t(\cos{k_x}+\cos{k_y})$ is the dispersion relation on the square lattice.

At large-$N$ limit, the saddle-point approximation of the partition function is accurate.
Since fermion operators of different flavors are decoupled in $S_E$, the mean-field equations, $\frac{\delta S_E[\chi,\vec{J}]}{\delta\chi}=\frac{\delta S_E[\chi,\vec{J}]}{\delta\vec{J}}=0$, can be simplified in the subspace of flavor $\alpha$ as
\begin{equation}\label{eq:mf-self}
    \begin{aligned}
        \frac{\chi_{ij}}{g_1} - \langle c_{i}^{\dagger}c_{j}+\mathrm{H.c.}\rangle_{h_E} &= 0,\\
        \frac{\vec{J}_{ij}}{g_2} - i\langle c_{i}^{\dagger}\frac{\vec{\sigma}}{2}c_{j}-\mathrm{H.c.}\rangle_{h_E} &= 0.
    \end{aligned}
\end{equation}


For the singlet $p_x$ density wave order, as in the right inset of Fig.~\ref{fig:fig1}, 
we obtain $\chi_{j,j+\delta} = \chi e^{iQ_xR_j}(\delta_{j+\delta,j+\hat{x}}-\delta_{j+\delta,j-\hat{x}})$.
Hence, the Fourier modes $\chi_{k'-k}$ are
\begin{equation}
    \begin{aligned}
        \chi_{k'-k} &= \frac{1}{2N_s}\sum_{j}\Big(\sum_{\delta}\chi_{j,j+\delta}e^{ik\delta}\Big)e^{-i(k'-k)R_j} \\
        &= i\chi\sin(k_x)\delta_{k',k+\vec{Q}_x}.
    \end{aligned}
\end{equation}
Substituting this into Eq.~\eqref{eq:effi_H}, we obtain the mean-field Hamiltonian matrix,
\begin{equation}\label{eq:h-mf-g1}
    h_E(\chi) = \sum_k
    \begin{pmatrix}
        \epsilon_k\sigma_0 & -i\chi\sin(k_x)\sigma_0  \\
        i\chi\sin(k_x)\sigma_0 & \epsilon_{k+\vec{Q}_x}\sigma_0
    \end{pmatrix},
\end{equation}
where the basis is $(c_{k\uparrow},c_{k\downarrow},c_{k+\vec{Q}_x\uparrow},c_{k+\vec{Q}_x\downarrow})$.
The dispersion relation of the singlet $p_x$ density wave state is $E_k(\chi)=-2t\cos{k_y}\pm\sqrt{\chi^2\sin^2{k_x}+4t^2\cos^2{k_x}}$. At $k_x=\pm\pi/2$, the energy gap between the upper and lower bands has a minimum value, $2\chi$. 
Similarly, consider the triplet $d_{x^2-y^2}$ density wave order as in the left inset of Fig.~\ref{fig:fig1}.
We have  $\vec{J}_{j,j+\delta} = -\frac{1}{4}\vec{J}e^{iQR_j}(\delta_{j+\delta,j+\hat{x}}+\delta_{j+\delta,j-\hat{x}}-\delta_{j+\delta,j+\hat{y}}-\delta_{j+\delta,j-\hat{y}})$.
The Fourier modes $\vec{J}_{k'-k}$ are
\begin{equation}
    \begin{aligned}
        \vec{J}_{k'-k} &= \frac{i}{2N_s}\sum_{j}\left(\sum_{\delta}\vec{J}_{j,j+\delta}e^{ik\delta}\right)e^{-i(k'-k)R_j} \\
        &= -\frac{i}{4}\vec{J}(\cos{k_x}-\cos{k_y})\delta_{k',k+\vec{Q}},
    \end{aligned}
\end{equation}
so the mean-field Hamiltonian matrix reduces to
\begin{equation}\label{eq:h-mf-g2}
    h_E(\vec{J}) = \sum_{k} 
    \begin{pmatrix}
        \epsilon_k\sigma_0 & -i\vec{\Delta}\cdot\frac{\vec{\sigma}}{2} \\
        i\vec{\Delta}\cdot\frac{\vec{\sigma}}{2} & \epsilon_{k+\vec{Q}}\sigma_0
    \end{pmatrix},
\end{equation}
with the basis $(c_{k\uparrow},c_{k\downarrow},c_{k+\vec{Q}\uparrow},c_{k+\vec{Q}\downarrow})$ and $\vec{\Delta}=-(\cos{k_x}-\cos{k_y})\vec{J}/4$. 
Then the dispersion relation is $E_k(\vec{J})=\pm\sqrt{\epsilon_k^2+(\vec{\Delta}/2)^2}$, which is a linear function around the Dirac points: $(\pm\pi/2,\pm\pi/2)$ and $(\mp\pi/2,\pm\pi/2)$.

\begin{figure}[tb]
  \includegraphics[width=0.8\linewidth]{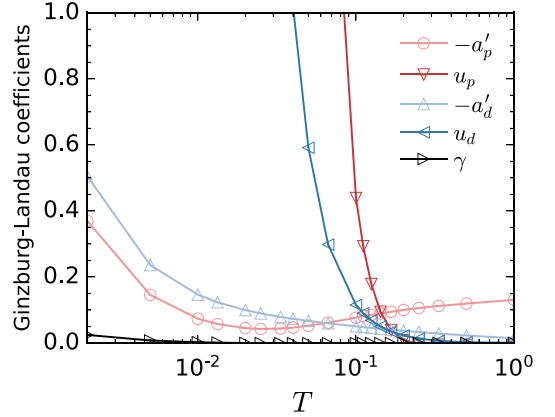}
  \caption{
          Coefficients in the Ginzburg-Landau free energy as a function of temperature. 
          The cutoff of Matsubara frequency $\omega_n$ with $n\leqslant20$ and lattice size $L=80$ are used.
  }\label{fig:supp:fig8}
\end{figure}

In the following, we calculate the Ginzburg-Landau (GL) free energy. We define the noninteracting Green's function $G_0(k)$ and mean-field operator $V_{k,k'}$ as $G_0(k) = (i\omega_n - \epsilon_k)^{-1}$ and $V_{k,k'}=-(\chi_{k'-k}\sigma_0+\vec{J}_{k'-k}\cdot\frac{\vec{\sigma}}{2})$, respectively. Therefore, the effective action can be written as
\begin{equation}
    \begin{aligned}
        &S_E[\chi,\vec{J}] = 2NN_s\beta \Big[\frac{\chi^2}{2g_1}+\frac{(\vec{J}/4)^2}{2g_2}\Big] \\
        &-N\mathrm{Tr}\ln[ (-i\omega_n + \epsilon_k)(1-G_0V_{k,k'})].        
    \end{aligned}
\end{equation}
For a noninteracting system, the free energy is given by
\begin{equation}
    F_0/N = \frac{S_0}{N_s\beta} = -\frac{1}{N_s\beta}\mathrm{Tr}\ln\left[ (-i\omega_n + \epsilon_k)\right].
\end{equation}
By expanding the remaining terms to the fourth order, we obtain the GL free energy,
\begin{equation}
        F/N = \frac{a_p}{2}\chi^2 + \frac{u_p}{4}\chi^4 + \frac{a_d}{2}\vec{J}^2 + \frac{u_d}{4}\vec{J}^4 + \frac{\gamma}{2}\chi^2 \vec{J}^2,
\end{equation}
where $a_p=\frac{2}{g_1}+a_{p}'$ and $a_d=\frac{1}{8g_2}+a_{d}'$. The coefficients $a'$, $u$, and $\gamma$ correspond to the Feynman diagrams
that can be solved numerically as a function of the temperature. 
As shown in Fig.~\ref{fig:supp:fig8}, the quadratic coefficients, $a'_p$ and $a'_d$, are negative, and they diverge while $T$ approaching the zero temperature limit. 
In contrast, the quartic coefficients $u_p$ and $u_d$ are positive.
Consequently, at zero temperature, there is a phase transition to the singlet $p_x$ (triplet $d_{x^2-y^2}$) density wave state at an infinitely small $g_1$ ($g_2$). 
Furthermore, $\gamma$ is positive, albeit small, and $\gamma<\sqrt{u_pu_d}$ in the low temperature regime.
A similar GL free energy was used to investigate the coexistence of SC and AFM orders \cite{fernandes2010competing,vorontsov2010super}.
Following the same line of Ref.~\cite{fernandes2010competing}, when the leading term for the description of competing orders satisfies $0<\gamma<\sqrt{u_pu_d}$, the two order parameters can be simultaneously nonzero.

\begin{figure}[t]
  \includegraphics[width=0.96\linewidth]{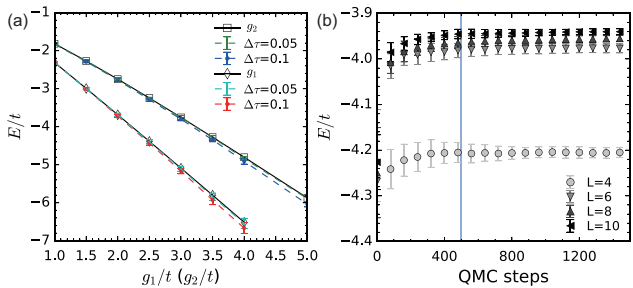}
  \caption{
          (a) Comparison between the PQMC and ED methods for finding the ground-state energy of the $N=1$ model on the $2\times2$ square lattice. 
          (b) Ground-state energy as a function of the QMC steps.
  }\label{fig:supp:fig9}
\end{figure}

\appsection{Error analysis}\label{sec:supp:qmc}
First, we employ the PQMC and exact diagonalization (ED) methods to find the ground-state energy $E$ of the $N=1$ model on the $2\times2$ square lattice. 
Figure~\ref{fig:supp:fig9}(a) shows representative data of $E$ versus $g_1$ ($g_2$) along the $g_1$ ($g_2$) axis for $\Delta\tau=0.1,0.05$. 
Though the deviations between the QMC and ED data for $\Delta\tau=0.1$ get bigger with increasing $g_1$ or $g_2$,
the PQMC method is still accurate within the error bars for the $N=1$ model at $g_{1,2}/t<5$. 
Nevertheless, the SU($N$)-symmetric Hamiltonian actually reduces $\Delta\tau$ by a factor of $N$, as shown in the HS decomposition~\eqref{eq:hs-decomp}. 
Therefore, $\Delta\tau=0.1$ should be sufficient for the parameter regimes used in our simulations.

Second, we determine the appropriate number of QMC steps for warming up and measurements. 
Different lattice sizes are considered because the number of auxiliary field rises when the number of lattice sites increases. 
In Fig.~\ref{fig:supp:fig9}(b), representative plots of $E$ versus QMC steps are presented for lattice sizes $L=4,6,8,10$.
From these data, we notice that the values of $E$ converge within the error bars 
after approximately 500 QMC steps. 
Therefore, we run $500$ QMC steps for warming up, followed by $500$ steps for measurements in each QMC bin.

\vspace{2cm}


\begin{thebibliography}{35}%
\makeatletter
\providecommand \@ifxundefined [1]{%
 \@ifx{#1\undefined}
}%
\providecommand \@ifnum [1]{%
 \ifnum #1\expandafter \@firstoftwo
 \else \expandafter \@secondoftwo
 \fi
}%
\providecommand \@ifx [1]{%
 \ifx #1\expandafter \@firstoftwo
 \else \expandafter \@secondoftwo
 \fi
}%
\providecommand \natexlab [1]{#1}%
\providecommand \enquote  [1]{``#1''}%
\providecommand \bibnamefont  [1]{#1}%
\providecommand \bibfnamefont [1]{#1}%
\providecommand \citenamefont [1]{#1}%
\providecommand \href@noop [0]{\@secondoftwo}%
\providecommand \href [0]{\begingroup \@sanitize@url \@href}%
\providecommand \@href[1]{\@@startlink{#1}\@@href}%
\providecommand \@@href[1]{\endgroup#1\@@endlink}%
\providecommand \@sanitize@url [0]{\catcode `\\12\catcode `\$12\catcode
  `\&12\catcode `\#12\catcode `\^12\catcode `\_12\catcode `\%12\relax}%
\providecommand \@@startlink[1]{}%
\providecommand \@@endlink[0]{}%
\providecommand \url  [0]{\begingroup\@sanitize@url \@url }%
\providecommand \@url [1]{\endgroup\@href {#1}{\urlprefix }}%
\providecommand \urlprefix  [0]{URL }%
\providecommand \Eprint [0]{\href }%
\providecommand \doibase [0]{http://dx.doi.org/}%
\providecommand \selectlanguage [0]{\@gobble}%
\providecommand \bibinfo  [0]{\@secondoftwo}%
\providecommand \bibfield  [0]{\@secondoftwo}%
\providecommand \translation [1]{[#1]}%
\providecommand \BibitemOpen [0]{}%
\providecommand \bibitemStop [0]{}%
\providecommand \bibitemNoStop [0]{.\EOS\space}%
\providecommand \EOS [0]{\spacefactor3000\relax}%
\providecommand \BibitemShut  [1]{\csname bibitem#1\endcsname}%
\let\auto@bib@innerbib\@empty
\bibitem [{\citenamefont {Nayak}(2000)}]{nayak2000density}%
  \BibitemOpen
  \bibfield  {author} {\bibinfo {author} {\bibfnamefont {C.}~\bibnamefont
  {Nayak}},\ }\href {\doibase 10.1103/PhysRevB.62.4880} {\bibfield  {journal}
  {\bibinfo  {journal} {Phys. Rev. B}\ }\textbf {\bibinfo {volume} {62}},\
  \bibinfo {pages} {4880} (\bibinfo {year} {2000})}\BibitemShut {NoStop}%
\bibitem [{\citenamefont {Lee}\ \emph {et~al.}(2006)\citenamefont {Lee},
  \citenamefont {Nagaosa},\ and\ \citenamefont {Wen}}]{lee2006doping}%
  \BibitemOpen
  \bibfield  {author} {\bibinfo {author} {\bibfnamefont {P.~A.}\ \bibnamefont
  {Lee}}, \bibinfo {author} {\bibfnamefont {N.}~\bibnamefont {Nagaosa}}, \ and\
  \bibinfo {author} {\bibfnamefont {X.-G.}\ \bibnamefont {Wen}},\ }\href
  {\doibase 10.1103/RevModPhys.78.17} {\bibfield  {journal} {\bibinfo
  {journal} {Rev. Mod. Phys.}\ }\textbf {\bibinfo {volume} {78}},\ \bibinfo
  {pages} {17} (\bibinfo {year} {2006})}\BibitemShut {NoStop}%
\bibitem [{\citenamefont {Affleck}\ and\ \citenamefont
  {Marston}(1988)}]{affleck1988largen}%
  \BibitemOpen
  \bibfield  {author} {\bibinfo {author} {\bibfnamefont {I.}~\bibnamefont
  {Affleck}}\ and\ \bibinfo {author} {\bibfnamefont {J.~B.}\ \bibnamefont
  {Marston}},\ }\href {\doibase 10.1103/PhysRevB.37.3774} {\bibfield  {journal}
  {\bibinfo  {journal} {Phys. Rev. B}\ }\textbf {\bibinfo {volume} {37}},\
  \bibinfo {pages} {3774} (\bibinfo {year} {1988})}\BibitemShut {NoStop}%
\bibitem [{\citenamefont {Marston}\ and\ \citenamefont
  {Affleck}(1989)}]{affleck1989largen}%
  \BibitemOpen
  \bibfield  {author} {\bibinfo {author} {\bibfnamefont {J.~B.}\ \bibnamefont
  {Marston}}\ and\ \bibinfo {author} {\bibfnamefont {I.}~\bibnamefont
  {Affleck}},\ }\href {\doibase 10.1103/PhysRevB.39.11538} {\bibfield
  {journal} {\bibinfo  {journal} {Phys. Rev. B}\ }\textbf {\bibinfo {volume}
  {39}},\ \bibinfo {pages} {11538} (\bibinfo {year} {1989})}\BibitemShut
  {NoStop}%
\bibitem [{\citenamefont {Wang}\ \emph {et~al.}(1990)\citenamefont {Wang},
  \citenamefont {Kotliar},\ and\ \citenamefont {Wang}}]{wang1990flux}%
  \BibitemOpen
  \bibfield  {author} {\bibinfo {author} {\bibfnamefont {Z.}~\bibnamefont
  {Wang}}, \bibinfo {author} {\bibfnamefont {G.}~\bibnamefont {Kotliar}}, \
  and\ \bibinfo {author} {\bibfnamefont {X.-F.}\ \bibnamefont {Wang}},\ }\href
  {\doibase 10.1103/PhysRevB.42.8690} {\bibfield  {journal} {\bibinfo
  {journal} {Phys. Rev. B}\ }\textbf {\bibinfo {volume} {42}},\ \bibinfo
  {pages} {8690} (\bibinfo {year} {1990})}\BibitemShut {NoStop}%
\bibitem [{\citenamefont {Liu}\ and\ \citenamefont
  {Wilczek}(2003)}]{liu2003spin}%
  \BibitemOpen
  \bibfield  {author} {\bibinfo {author} {\bibfnamefont {W.~V.}\ \bibnamefont
  {Liu}}\ and\ \bibinfo {author} {\bibfnamefont {F.}~\bibnamefont {Wilczek}},\
  }\href {https://doi.org/10.48550/arXiv.cond-mat/0312685} {\bibfield
  {journal} {\bibinfo  {journal} {arXiv preprint cond-mat/0312685}\ } (\bibinfo
  {year} {2003})}\BibitemShut {NoStop}%
\bibitem [{\citenamefont {Chakravarty}\ \emph {et~al.}(2001)\citenamefont
  {Chakravarty}, \citenamefont {Laughlin}, \citenamefont {Morr},\ and\
  \citenamefont {Nayak}}]{chakravarty2001hidden}%
  \BibitemOpen
  \bibfield  {author} {\bibinfo {author} {\bibfnamefont {S.}~\bibnamefont
  {Chakravarty}}, \bibinfo {author} {\bibfnamefont {R.~B.}\ \bibnamefont
  {Laughlin}}, \bibinfo {author} {\bibfnamefont {D.~K.}\ \bibnamefont {Morr}},
  \ and\ \bibinfo {author} {\bibfnamefont {C.}~\bibnamefont {Nayak}},\ }\href
  {\doibase 10.1103/PhysRevB.63.094503} {\bibfield  {journal} {\bibinfo
  {journal} {Phys. Rev. B}\ }\textbf {\bibinfo {volume} {63}},\ \bibinfo
  {pages} {094503} (\bibinfo {year} {2001})}\BibitemShut {NoStop}%
\bibitem [{\citenamefont {Maki}\ \emph {et~al.}(2007)\citenamefont {Maki},
  \citenamefont {D{\'o}ra}, \citenamefont {V{\'a}nyolos},\ and\ \citenamefont
  {Virosztek}}]{maki2007d}%
  \BibitemOpen
  \bibfield  {author} {\bibinfo {author} {\bibfnamefont {K.}~\bibnamefont
  {Maki}}, \bibinfo {author} {\bibfnamefont {B.}~\bibnamefont {D{\'o}ra}},
  \bibinfo {author} {\bibfnamefont {A.}~\bibnamefont {V{\'a}nyolos}}, \ and\
  \bibinfo {author} {\bibfnamefont {A.}~\bibnamefont {Virosztek}},\ }\href
  {\doibase 10.1016/j.physc.2007.03.014} {\bibfield  {journal} {\bibinfo
  {journal} {Physica C: Supercond.}\ }\textbf {\bibinfo {volume} {460-462}},\
  \bibinfo {pages} {226} (\bibinfo {year} {2007})}\BibitemShut {NoStop}%
\bibitem [{\citenamefont {Maharaj}\ \emph {et~al.}(2013)\citenamefont
  {Maharaj}, \citenamefont {Thomale},\ and\ \citenamefont
  {Raghu}}]{maharaj2013particle}%
  \BibitemOpen
  \bibfield  {author} {\bibinfo {author} {\bibfnamefont {A.~V.}\ \bibnamefont
  {Maharaj}}, \bibinfo {author} {\bibfnamefont {R.}~\bibnamefont {Thomale}}, \
  and\ \bibinfo {author} {\bibfnamefont {S.}~\bibnamefont {Raghu}},\ }\href
  {\doibase 10.1103/PhysRevB.88.205121} {\bibfield  {journal} {\bibinfo
  {journal} {Phys. Rev. B}\ }\textbf {\bibinfo {volume} {88}},\ \bibinfo
  {pages} {205121} (\bibinfo {year} {2013})}\BibitemShut {NoStop}%
\bibitem [{\citenamefont
  {Venderbos}(2016{\natexlab{a}})}]{venderbos2016symmetry}%
  \BibitemOpen
  \bibfield  {author} {\bibinfo {author} {\bibfnamefont {J.~W.~F.}\
  \bibnamefont {Venderbos}},\ }\href {\doibase 10.1103/PhysRevB.93.115107}
  {\bibfield  {journal} {\bibinfo  {journal} {Phys. Rev. B}\ }\textbf {\bibinfo
  {volume} {93}},\ \bibinfo {pages} {115107} (\bibinfo {year}
  {2016}{\natexlab{a}})}\BibitemShut {NoStop}%
\bibitem [{\citenamefont {Venderbos}(2016{\natexlab{b}})}]{venderbos2016multi}%
  \BibitemOpen
  \bibfield  {author} {\bibinfo {author} {\bibfnamefont {J.~W.~F.}\
  \bibnamefont {Venderbos}},\ }\href {\doibase 10.1103/PhysRevB.93.115108}
  {\bibfield  {journal} {\bibinfo  {journal} {Phys. Rev. B}\ }\textbf {\bibinfo
  {volume} {93}},\ \bibinfo {pages} {115108} (\bibinfo {year}
  {2016}{\natexlab{b}})}\BibitemShut {NoStop}%
\bibitem [{\citenamefont {Assaad}(2005)}]{assaad2005phase}%
  \BibitemOpen
  \bibfield  {author} {\bibinfo {author} {\bibfnamefont {F.~F.}\ \bibnamefont
  {Assaad}},\ }\href {\doibase 10.1103/PhysRevB.71.075103} {\bibfield
  {journal} {\bibinfo  {journal} {Phys. Rev. B}\ }\textbf {\bibinfo {volume}
  {71}},\ \bibinfo {pages} {075103} (\bibinfo {year} {2005})}\BibitemShut
  {NoStop}%
\bibitem [{\citenamefont {Lang}\ \emph {et~al.}(2013)\citenamefont {Lang},
  \citenamefont {Meng}, \citenamefont {Muramatsu}, \citenamefont {Wessel},\
  and\ \citenamefont {Assaad}}]{lang2013dimerized}%
  \BibitemOpen
  \bibfield  {author} {\bibinfo {author} {\bibfnamefont {T.~C.}\ \bibnamefont
  {Lang}}, \bibinfo {author} {\bibfnamefont {Z.~Y.}\ \bibnamefont {Meng}},
  \bibinfo {author} {\bibfnamefont {A.}~\bibnamefont {Muramatsu}}, \bibinfo
  {author} {\bibfnamefont {S.}~\bibnamefont {Wessel}}, \ and\ \bibinfo {author}
  {\bibfnamefont {F.~F.}\ \bibnamefont {Assaad}},\ }\href {\doibase
  10.1103/PhysRevLett.111.066401} {\bibfield  {journal} {\bibinfo  {journal}
  {Phys. Rev. Lett.}\ }\textbf {\bibinfo {volume} {111}},\ \bibinfo {pages}
  {066401} (\bibinfo {year} {2013})}\BibitemShut {NoStop}%
\bibitem [{\citenamefont {Wu}\ \emph {et~al.}(2003)\citenamefont {Wu},
  \citenamefont {Hu},\ and\ \citenamefont {Zhang}}]{Wu2003Exact}%
  \BibitemOpen
  \bibfield  {author} {\bibinfo {author} {\bibfnamefont {C.}~\bibnamefont
  {Wu}}, \bibinfo {author} {\bibfnamefont {J.-p.}\ \bibnamefont {Hu}}, \ and\
  \bibinfo {author} {\bibfnamefont {S.-c.}\ \bibnamefont {Zhang}},\ }\href
  {\doibase 10.1103/PhysRevLett.91.186402} {\bibfield  {journal} {\bibinfo
  {journal} {Phys. Rev. Lett.}\ }\textbf {\bibinfo {volume} {91}},\ \bibinfo
  {pages} {186402} (\bibinfo {year} {2003})}\BibitemShut {NoStop}%
\bibitem [{\citenamefont {Wu}(2006)}]{wu2006hidden}%
  \BibitemOpen
  \bibfield  {author} {\bibinfo {author} {\bibfnamefont {C.}~\bibnamefont
  {Wu}},\ }\href {\doibase 10.1142/S0217984906012213} {\bibfield  {journal}
  {\bibinfo  {journal} {Mod. Phys. Lett. B}\ }\textbf {\bibinfo {volume}
  {20}},\ \bibinfo {pages} {1707} (\bibinfo {year} {2006})}\BibitemShut
  {NoStop}%
\bibitem [{\citenamefont {DeSalvo}\ \emph {et~al.}(2010)\citenamefont
  {DeSalvo}, \citenamefont {Yan}, \citenamefont {Mickelson}, \citenamefont
  {Martinez~de Escobar},\ and\ \citenamefont
  {Killian}}]{DeSalvo2010Degenerate}%
  \BibitemOpen
  \bibfield  {author} {\bibinfo {author} {\bibfnamefont {B.~J.}\ \bibnamefont
  {DeSalvo}}, \bibinfo {author} {\bibfnamefont {M.}~\bibnamefont {Yan}},
  \bibinfo {author} {\bibfnamefont {P.~G.}\ \bibnamefont {Mickelson}}, \bibinfo
  {author} {\bibfnamefont {Y.~N.}\ \bibnamefont {Martinez~de Escobar}}, \ and\
  \bibinfo {author} {\bibfnamefont {T.~C.}\ \bibnamefont {Killian}},\ }\href
  {\doibase 10.1103/PhysRevLett.105.030402} {\bibfield  {journal} {\bibinfo
  {journal} {Phys. Rev. Lett.}\ }\textbf {\bibinfo {volume} {105}},\ \bibinfo
  {pages} {030402} (\bibinfo {year} {2010})}\BibitemShut {NoStop}%
\bibitem [{\citenamefont {Zhang}\ \emph {et~al.}(2014)\citenamefont {Zhang},
  \citenamefont {Bishof}, \citenamefont {Bromley}, \citenamefont {Kraus},
  \citenamefont {Safronova}, \citenamefont {Zoller}, \citenamefont {Rey},\ and\
  \citenamefont {Ye}}]{zhang2014spectroscopic}%
  \BibitemOpen
  \bibfield  {author} {\bibinfo {author} {\bibfnamefont {X.}~\bibnamefont
  {Zhang}}, \bibinfo {author} {\bibfnamefont {M.}~\bibnamefont {Bishof}},
  \bibinfo {author} {\bibfnamefont {S.~L.}\ \bibnamefont {Bromley}}, \bibinfo
  {author} {\bibfnamefont {C.~V.}\ \bibnamefont {Kraus}}, \bibinfo {author}
  {\bibfnamefont {M.~S.}\ \bibnamefont {Safronova}}, \bibinfo {author}
  {\bibfnamefont {P.}~\bibnamefont {Zoller}}, \bibinfo {author} {\bibfnamefont
  {A.~M.}\ \bibnamefont {Rey}}, \ and\ \bibinfo {author} {\bibfnamefont
  {J.}~\bibnamefont {Ye}},\ }\href {\doibase 10.1126/science.1254978}
  {\bibfield  {journal} {\bibinfo  {journal} {Science}\ }\textbf {\bibinfo
  {volume} {345}},\ \bibinfo {pages} {1467} (\bibinfo {year}
  {2014})}\BibitemShut {NoStop}%
\bibitem [{\citenamefont {Taie}\ \emph {et~al.}(2010)\citenamefont {Taie},
  \citenamefont {Takasu}, \citenamefont {Sugawa}, \citenamefont {Yamazaki},
  \citenamefont {Tsujimoto}, \citenamefont {Murakami},\ and\ \citenamefont
  {Takahashi}}]{taie2010realization}%
  \BibitemOpen
  \bibfield  {author} {\bibinfo {author} {\bibfnamefont {S.}~\bibnamefont
  {Taie}}, \bibinfo {author} {\bibfnamefont {Y.}~\bibnamefont {Takasu}},
  \bibinfo {author} {\bibfnamefont {S.}~\bibnamefont {Sugawa}}, \bibinfo
  {author} {\bibfnamefont {R.}~\bibnamefont {Yamazaki}}, \bibinfo {author}
  {\bibfnamefont {T.}~\bibnamefont {Tsujimoto}}, \bibinfo {author}
  {\bibfnamefont {R.}~\bibnamefont {Murakami}}, \ and\ \bibinfo {author}
  {\bibfnamefont {Y.}~\bibnamefont {Takahashi}},\ }\href {\doibase
  10.1103/PhysRevLett.105.190401} {\bibfield  {journal} {\bibinfo  {journal}
  {Phys. Rev. Lett.}\ }\textbf {\bibinfo {volume} {105}},\ \bibinfo {pages}
  {190401} (\bibinfo {year} {2010})}\BibitemShut {NoStop}%
\bibitem [{\citenamefont {Taie}\ \emph {et~al.}(2012)\citenamefont {Taie},
  \citenamefont {Yamazaki}, \citenamefont {Sugawa},\ and\ \citenamefont
  {Takahashi}}]{taie2012ansu6}%
  \BibitemOpen
  \bibfield  {author} {\bibinfo {author} {\bibfnamefont {S.}~\bibnamefont
  {Taie}}, \bibinfo {author} {\bibfnamefont {R.}~\bibnamefont {Yamazaki}},
  \bibinfo {author} {\bibfnamefont {S.}~\bibnamefont {Sugawa}}, \ and\ \bibinfo
  {author} {\bibfnamefont {Y.}~\bibnamefont {Takahashi}},\ }\href {\doibase
  10.1038/nphys2430} {\bibfield  {journal} {\bibinfo  {journal} {Nat. Phys.}\
  }\textbf {\bibinfo {volume} {8}},\ \bibinfo {pages} {825} (\bibinfo {year}
  {2012})}\BibitemShut {NoStop}%
\bibitem [{\citenamefont {Taie}\ \emph {et~al.}(2022)\citenamefont {Taie},
  \citenamefont {Ibarra-Garc{\'\i}a-Padilla}, \citenamefont {Nishizawa},
  \citenamefont {Takasu}, \citenamefont {Kuno}, \citenamefont {Wei},
  \citenamefont {Scalettar}, \citenamefont {Hazzard},\ and\ \citenamefont
  {Takahashi}}]{taie2022observation}%
  \BibitemOpen
  \bibfield  {author} {\bibinfo {author} {\bibfnamefont {S.}~\bibnamefont
  {Taie}}, \bibinfo {author} {\bibfnamefont {E.}~\bibnamefont
  {Ibarra-Garc{\'\i}a-Padilla}}, \bibinfo {author} {\bibfnamefont
  {N.}~\bibnamefont {Nishizawa}}, \bibinfo {author} {\bibfnamefont
  {Y.}~\bibnamefont {Takasu}}, \bibinfo {author} {\bibfnamefont
  {Y.}~\bibnamefont {Kuno}}, \bibinfo {author} {\bibfnamefont {H.-T.}\
  \bibnamefont {Wei}}, \bibinfo {author} {\bibfnamefont {R.~T.}\ \bibnamefont
  {Scalettar}}, \bibinfo {author} {\bibfnamefont {K.~R.}\ \bibnamefont
  {Hazzard}}, \ and\ \bibinfo {author} {\bibfnamefont {Y.}~\bibnamefont
  {Takahashi}},\ }\href {https://doi.org/10.1038/s41567-022-01725-6} {\bibfield
   {journal} {\bibinfo  {journal} {Nat. Phys.}\ } (\bibinfo {year}
  {2022})}\BibitemShut {NoStop}%
\bibitem [{\citenamefont {Li}\ \emph {et~al.}(2017)\citenamefont {Li},
  \citenamefont {Jiang}, \citenamefont {Jian},\ and\ \citenamefont
  {Yao}}]{li2017fermion}%
  \BibitemOpen
  \bibfield  {author} {\bibinfo {author} {\bibfnamefont {Z.-X.}\ \bibnamefont
  {Li}}, \bibinfo {author} {\bibfnamefont {Y.-F.}\ \bibnamefont {Jiang}},
  \bibinfo {author} {\bibfnamefont {S.-K.}\ \bibnamefont {Jian}}, \ and\
  \bibinfo {author} {\bibfnamefont {H.}~\bibnamefont {Yao}},\ }\href {\doibase
  10.1038/s41467-017-00167-6} {\bibfield  {journal} {\bibinfo  {journal} {Nat.
  Commun.}\ }\textbf {\bibinfo {volume} {8}},\ \bibinfo {pages} {1} (\bibinfo
  {year} {2017})}\BibitemShut {NoStop}%
\bibitem [{\citenamefont {Capponi}\ and\ \citenamefont
  {Assaad}(2007)}]{capponi2007spin}%
  \BibitemOpen
  \bibfield  {author} {\bibinfo {author} {\bibfnamefont {S.}~\bibnamefont
  {Capponi}}\ and\ \bibinfo {author} {\bibfnamefont {F.~F.}\ \bibnamefont
  {Assaad}},\ }\href {\doibase 10.1103/PhysRevB.75.045115} {\bibfield
  {journal} {\bibinfo  {journal} {Phys. Rev. B}\ }\textbf {\bibinfo {volume}
  {75}},\ \bibinfo {pages} {045115} (\bibinfo {year} {2007})}\BibitemShut
  {NoStop}%
\bibitem [{\citenamefont {Wu}\ and\ \citenamefont
  {Zhang}(2005)}]{wu2005sufficient}%
  \BibitemOpen
  \bibfield  {author} {\bibinfo {author} {\bibfnamefont {C.}~\bibnamefont
  {Wu}}\ and\ \bibinfo {author} {\bibfnamefont {S.-C.}\ \bibnamefont {Zhang}},\
  }\href {\doibase 10.1103/PhysRevB.71.155115} {\bibfield  {journal} {\bibinfo
  {journal} {Phys. Rev. B}\ }\textbf {\bibinfo {volume} {71}},\ \bibinfo
  {pages} {155115} (\bibinfo {year} {2005})}\BibitemShut {NoStop}%
\bibitem [{\citenamefont {Blankenbecler}\ \emph {et~al.}(1981)\citenamefont
  {Blankenbecler}, \citenamefont {Scalapino},\ and\ \citenamefont
  {Sugar}}]{blankenbecler1981monte}%
  \BibitemOpen
  \bibfield  {author} {\bibinfo {author} {\bibfnamefont {R.}~\bibnamefont
  {Blankenbecler}}, \bibinfo {author} {\bibfnamefont {D.~J.}\ \bibnamefont
  {Scalapino}}, \ and\ \bibinfo {author} {\bibfnamefont {R.~L.}\ \bibnamefont
  {Sugar}},\ }\href {\doibase 10.1103/PhysRevD.24.2278} {\bibfield  {journal}
  {\bibinfo  {journal} {Phys. Rev. D}\ }\textbf {\bibinfo {volume} {24}},\
  \bibinfo {pages} {2278} (\bibinfo {year} {1981})}\BibitemShut {NoStop}%
\bibitem [{\citenamefont {Hirsch}(1985)}]{hirsh1985two}%
  \BibitemOpen
  \bibfield  {author} {\bibinfo {author} {\bibfnamefont {J.~E.}\ \bibnamefont
  {Hirsch}},\ }\href {\doibase 10.1103/PhysRevB.31.4403} {\bibfield  {journal}
  {\bibinfo  {journal} {Phys. Rev. B}\ }\textbf {\bibinfo {volume} {31}},\
  \bibinfo {pages} {4403} (\bibinfo {year} {1985})}\BibitemShut {NoStop}%
\bibitem [{\citenamefont {Assaad}\ and\ \citenamefont
  {Evertz}(2008)}]{Assaad2008Worldline}%
  \BibitemOpen
  \bibfield  {author} {\bibinfo {author} {\bibfnamefont {F.}~\bibnamefont
  {Assaad}}\ and\ \bibinfo {author} {\bibfnamefont {H.}~\bibnamefont
  {Evertz}},\ }\enquote {\bibinfo {title} {World-line and determinantal quantum
  {M}onte {C}arlo methods for spins, phonons and electrons},}\ in\ \href
  {\doibase 10.1007/978-3-540-74686-7_10} {\emph {\bibinfo {booktitle}
  {Computational Many-Particle Physics}}}\ (\bibinfo  {publisher} {Springer
  Berlin Heidelberg},\ \bibinfo {year} {2008})\ pp.\ \bibinfo {pages}
  {277--356}\BibitemShut {NoStop}%
\bibitem [{\citenamefont {Wang}\ \emph {et~al.}(2014)\citenamefont {Wang},
  \citenamefont {Li}, \citenamefont {Cai}, \citenamefont {Zhou}, \citenamefont
  {Wang},\ and\ \citenamefont {Wu}}]{wang2014competing}%
  \BibitemOpen
  \bibfield  {author} {\bibinfo {author} {\bibfnamefont {D.}~\bibnamefont
  {Wang}}, \bibinfo {author} {\bibfnamefont {Y.}~\bibnamefont {Li}}, \bibinfo
  {author} {\bibfnamefont {Z.}~\bibnamefont {Cai}}, \bibinfo {author}
  {\bibfnamefont {Z.}~\bibnamefont {Zhou}}, \bibinfo {author} {\bibfnamefont
  {Y.}~\bibnamefont {Wang}}, \ and\ \bibinfo {author} {\bibfnamefont
  {C.}~\bibnamefont {Wu}},\ }\href {\doibase 10.1103/PhysRevLett.112.156403}
  {\bibfield  {journal} {\bibinfo  {journal} {Phys. Rev. Lett.}\ }\textbf
  {\bibinfo {volume} {112}},\ \bibinfo {pages} {156403} (\bibinfo {year}
  {2014})}\BibitemShut {NoStop}%
\bibitem [{SourceCode()}]{SourceCode}%
  \BibitemOpen
  \bibinfo {note} {Source codes used in this paper
  are available at
  https://github.com/MilCOS/QMC-for-bond-and-current-interactions}\BibitemShut
  {NoStop}%
\bibitem [{\citenamefont {Watanabe}\ and\ \citenamefont
  {Usui}(1985)}]{watanabe1985phase}%
  \BibitemOpen
  \bibfield  {author} {\bibinfo {author} {\bibfnamefont {S.}~\bibnamefont
  {Watanabe}}\ and\ \bibinfo {author} {\bibfnamefont {T.}~\bibnamefont
  {Usui}},\ }\href {\doibase 10.1143/PTP.73.1305} {\bibfield  {journal}
  {\bibinfo  {journal} {Prog. Theor. Phys.}\ }\textbf {\bibinfo {volume}
  {73}},\ \bibinfo {pages} {1305} (\bibinfo {year} {1985})}\BibitemShut
  {NoStop}%
\bibitem [{\citenamefont {Anisimov}\ \emph {et~al.}(1981)\citenamefont
  {Anisimov}, \citenamefont {Gorodetski{\u{\i}}},\ and\ \citenamefont
  {Zaprudski{\u{\i}}}}]{anisimov1981phase}%
  \BibitemOpen
  \bibfield  {author} {\bibinfo {author} {\bibfnamefont {M.~A.}\ \bibnamefont
  {Anisimov}}, \bibinfo {author} {\bibfnamefont {E.~E.}\ \bibnamefont
  {Gorodetski{\u{\i}}}}, \ and\ \bibinfo {author} {\bibfnamefont {V.~M.}\
  \bibnamefont {Zaprudski{\u{\i}}}},\ }\href {\doibase
  10.1070/pu1981v024n01abeh004612} {\bibfield  {journal} {\bibinfo  {journal}
  {Sov. Phys. Usp.}\ }\textbf {\bibinfo {volume} {24}},\ \bibinfo {pages} {57}
  (\bibinfo {year} {1981})}\BibitemShut {NoStop}%
\bibitem [{\citenamefont {Parisen~Toldin}\ \emph {et~al.}(2015)\citenamefont
  {Parisen~Toldin}, \citenamefont {Hohenadler}, \citenamefont {Assaad},\ and\
  \citenamefont {Herbut}}]{Parisen2015Fermionic}%
  \BibitemOpen
  \bibfield  {author} {\bibinfo {author} {\bibfnamefont {F.}~\bibnamefont
  {Parisen~Toldin}}, \bibinfo {author} {\bibfnamefont {M.}~\bibnamefont
  {Hohenadler}}, \bibinfo {author} {\bibfnamefont {F.~F.}\ \bibnamefont
  {Assaad}}, \ and\ \bibinfo {author} {\bibfnamefont {I.~F.}\ \bibnamefont
  {Herbut}},\ }\href {\doibase 10.1103/PhysRevB.91.165108} {\bibfield
  {journal} {\bibinfo  {journal} {Phys. Rev. B}\ }\textbf {\bibinfo {volume}
  {91}},\ \bibinfo {pages} {165108} (\bibinfo {year} {2015})}\BibitemShut
  {NoStop}%
\bibitem [{\citenamefont {Otsuki}\ \emph {et~al.}(2017)\citenamefont {Otsuki},
  \citenamefont {Ohzeki}, \citenamefont {Shinaoka},\ and\ \citenamefont
  {Yoshimi}}]{otsuki2017sparse}%
  \BibitemOpen
  \bibfield  {author} {\bibinfo {author} {\bibfnamefont {J.}~\bibnamefont
  {Otsuki}}, \bibinfo {author} {\bibfnamefont {M.}~\bibnamefont {Ohzeki}},
  \bibinfo {author} {\bibfnamefont {H.}~\bibnamefont {Shinaoka}}, \ and\
  \bibinfo {author} {\bibfnamefont {K.}~\bibnamefont {Yoshimi}},\ }\href
  {\doibase 10.1103/PhysRevE.95.061302} {\bibfield  {journal} {\bibinfo
  {journal} {Phys. Rev. E}\ }\textbf {\bibinfo {volume} {95}},\ \bibinfo
  {pages} {061302} (\bibinfo {year} {2017})}\BibitemShut {NoStop}%
\bibitem [{\citenamefont {Coleman}(2015)}]{coleman2015introduction}%
  \BibitemOpen
  \bibfield  {author} {\bibinfo {author} {\bibfnamefont {P.}~\bibnamefont
  {Coleman}},\ }\href@noop {} {\emph {\bibinfo {title} {Introduction to
  many-body physics}}}\ (\bibinfo  {publisher} {Cambridge University Press},\
  \bibinfo {year} {2015})\BibitemShut {NoStop}%
\bibitem [{\citenamefont {Fernandes}\ and\ \citenamefont
  {Schmalian}(2010)}]{fernandes2010competing}%
  \BibitemOpen
  \bibfield  {author} {\bibinfo {author} {\bibfnamefont {R.~M.}\ \bibnamefont
  {Fernandes}}\ and\ \bibinfo {author} {\bibfnamefont {J.}~\bibnamefont
  {Schmalian}},\ }\href {\doibase 10.1103/PhysRevB.82.014521} {\bibfield
  {journal} {\bibinfo  {journal} {Phys. Rev. B}\ }\textbf {\bibinfo {volume}
  {82}},\ \bibinfo {pages} {014521} (\bibinfo {year} {2010})}\BibitemShut
  {NoStop}%
\bibitem [{\citenamefont {Vorontsov}\ \emph {et~al.}(2010)\citenamefont
  {Vorontsov}, \citenamefont {Vavilov},\ and\ \citenamefont
  {Chubukov}}]{vorontsov2010super}%
  \BibitemOpen
  \bibfield  {author} {\bibinfo {author} {\bibfnamefont {A.~B.}\ \bibnamefont
  {Vorontsov}}, \bibinfo {author} {\bibfnamefont {M.~G.}\ \bibnamefont
  {Vavilov}}, \ and\ \bibinfo {author} {\bibfnamefont {A.~V.}\ \bibnamefont
  {Chubukov}},\ }\href {\doibase 10.1103/PhysRevB.81.174538} {\bibfield
  {journal} {\bibinfo  {journal} {Phys. Rev. B}\ }\textbf {\bibinfo {volume}
  {81}},\ \bibinfo {pages} {174538} (\bibinfo {year} {2010})}\BibitemShut
  {NoStop}%
\end{thebibliography}
%

\end{document}